\documentclass[twoside,magazine]{IEEEtran}
\usepackage{amsmath}
\usepackage{amsfonts,amssymb}
\usepackage{makecell}
\usepackage{hyperref}
\usepackage{multirow}
\usepackage{array}
\usepackage{graphicx,amssymb,amsmath}
\usepackage{multicol}
\usepackage[noadjust]{cite}
\usepackage{setspace}
\usepackage{subfig}
\usepackage{graphicx}
\usepackage{float}
\usepackage {url}
\usepackage{stfloats}
\usepackage{amsthm,pifont}
\usepackage{flushend}
\usepackage{cases,subeqnarray}
\usepackage{bm,multirow,bigstrut}
\usepackage{amsmath, amsthm, amssymb}
\usepackage{textcomp}
\usepackage{latexsym,bm}
\usepackage{booktabs}
\usepackage{xcolor}
\usepackage{mathtools}
\usepackage{dsfont}
\usepackage{extarrows}
\usepackage{epsfig}
\usepackage{epsfig}
\usepackage{epstopdf}
\usepackage[noend]{algpseudocode}
\usepackage{algorithmicx,algorithm}
\usepackage{amssymb}
\usepackage{xspace}
\usepackage{listings}
\usepackage{xcolor}
\definecolor{commentcolor}{RGB}{85,139,78}
\definecolor{stringcolor}{RGB}{206,145,108}
\definecolor{keywordcolor}{RGB}{0,0,128}
\definecolor{backcolor}{RGB}{220,220,220}
\theoremstyle{plain}

\theoremstyle{plain}

\usepackage{amsmath}

\IEEEoverridecommandlockouts

\begin{document}
\title{A Unified Framework for Guiding Generative AI with Wireless Perception in Resource Constrained Mobile Edge Networks}
\author{Jiacheng Wang, Hongyang Du, Dusit~Niyato,~\IEEEmembership{Fellow,~IEEE}, Jiawen Kang, Zehui Xiong, Deepu Rajan, \\ Shiwen Mao,~\IEEEmembership{Fellow,~IEEE}, and Xuemin~(Sherman)~Shen,~\IEEEmembership{Fellow,~IEEE}
 \thanks{J.~Wang, H.~Du, D.~Niyato, and D.~Rajan are with the School of Computer Science and Engineering, Nanyang Technological University, Singapore (e-mail: jiacheng.wang@ntu.edu.sg, hongyang001@e.ntu.edu.sg, dniyato@ntu.edu.sg, asdrajan@ntu.edu.sg).}
\thanks{J. Kang is with the School of Automation, Guangdong University of Technology, China (e-mail: kavinkang@gdut.edu.cn).}
\thanks{Z. Xiong is with the Pillar of Information Systems Technology and Design, Singapore University of Technology and Design, Singapore (e-mail: zehui\_xiong@sutd.edu.sg).}
 
 \thanks{S. Mao is with the Department of Electrical and Computer Engineering, Auburn University, Auburn, USA (e-mail: smao@ieee.org)}
 \thanks{X. Shen is with the Department of Electrical and Computer Engineering, University of Waterloo, Canada (e-mail: sshen@uwaterloo.ca).}

}

\maketitle
\begin{abstract}
With the significant advancements in artificial intelligence (AI) technologies and powerful computational capabilities, generative AI (GAI) has become a pivotal digital content generation technique for offering superior digital services. However, directing GAI towards desired outputs still suffer the inherent instability of the AI model. In this paper, we design a novel framework that utilizes \underline{wi}reless \underline{pe}rception to guide \underline{GAI} (WiPe-GAI) for providing digital content generation service, i.e., AI-generated content (AIGC), in resource-constrained mobile edge networks. Specifically, we first propose a new sequential multi-scale perception (SMSP) algorithm to predict user skeleton based on the channel state information (CSI) extracted from wireless signals. This prediction then guides GAI to provide users with AIGC, such as virtual character generation. To ensure the efficient operation of the proposed framework in resource constrained networks, we further design a pricing-based incentive mechanism and introduce a diffusion model based approach to generate an optimal pricing strategy for the service provisioning. The strategy maximizes the user’s utility while enhancing the participation of the virtual service provider (VSP) in AIGC provision. The experimental results demonstrate the effectiveness of the designed framework in terms of skeleton prediction and optimal pricing strategy generation comparing with other existing solutions.
\end{abstract}

\IEEEpeerreviewmaketitle
\section{Introduction}
In recent years, the accelerated proliferation of diverse user data, advancements in hardware devices, and the evolution of AI models catalyze the rapid progression of generative artificial intelligence (GAI) technology~\cite{xu2023unleashing}. As a result, the artificial intelligence-generated content (AIGC) and its associated applications attract considerable attention~\cite{koksal2023controllable}. Major technological giants, such as Microsoft and Google, invest heavily in creating their own exclusive GAI model, with the objective of offering users a more comprehensive digital service~\cite{wu2023ai}. A representative work is OpenAI's ChatGPT, which achieves notable breakthroughs in emulating human in text processing tasks. For instance, ChatGPT is capable of not only executing grammar error detection and refinement, but also generating text, code, and performing content retrieval operations~\cite{cao2023comprehensive}. Beyond text processing, the powerful capabilities of GAI are also unleashed in the realm of image and video generation. For instance, Stable Diffusion can generate images based on users' descriptions (i.e., prompts), as well as process images according to users' instructions, including style modifications and rectification of missing pixels and other visual imperfections~\cite{croitoru2023diffusion}.

In comparison to the conventional generation methods, GAI exhibits two salient advantages. First, GAI has a superior productivity, capable of generating digital content quickly in accordance with user directives. For example, stable diffusion model~\cite{zhang2023adding} can generate a high-definition image within seconds, which is challenging to accomplish by a traditional user based generation method. Second, AIGC exhibits greater diversity, manifested in two aspects~\cite{wang2023guiding}. The first aspect pertains to the richness of the generated content. Owing to the randomness of the seed in AI models, GAI's outputs can vary significantly even with identical instructions. For example, the diffusion model can generate entirely different images with the same input prompt, thus offering users a broader range of choices. The second aspect is the multimodal presentation format, which allows AIGC to be delivered in various forms such as text, images, videos, and even audio~\cite{bond2021deep}. This makes AIGC highly adaptable, catering to a range of applications. Due to these aforementioned benefits, GAI has emerged as the critical engine for creating digital content, playing an indispensable role in our progression towards a more immersive and interactive next-generation Internet~\cite{du2023generative}.

Despite the significant advancements, several challenges still need to be tackled for practical applications. \textbf{\emph {First, the inherent instability of AI models makes it difficult to meet users' needs, especially when generating digital content directly related to users themselves}}~\cite{zhang2023adding}. For example, in augmented reality (AR) applications, such as virtual game and shopping, the virtual service providers (VSPs) use the GAI technology to create virtual characters for users. However, due to the randomness of seeds in AI model and the difficulty of conveying information through prompts about users' posture to the AI model, the generated characters may not align accurately with the actual user. As a result, users may generate multiple requests until a satisfactory character is generated, which not only degrades the quality of service (QoS) of the VSP, but also leads to resource wastage. To mitigate this, an effective solution is to guide GAI with the help of other methods. However, the computation resources of the VSP deployed in mobile edge networks are typically limited. \textbf{\emph {This leads to the second challenge when employing other methods to guide GAI, that is, how to incentivize the VSPs to engage actively in service provision, thereby ensuring the efficient operation of the overall framework.}} A potential solution to this issue entails establishing a payment plan between the user and the VSP, whereby the user provides fee to the VSP according to the plan to encourage participation.

Given the aforementioned challenges and potential solutions, this paper introduces wireless perception guided GAI (WiPe-GAI), a novel framework deployed in resource-constrained mobile edge networks, which uses wireless perception to guide GAI in providing AIGC to users and introduces an incentive mechanism to ensure its economical operation. Specifically, in WiPe-GAI, we first propose a novel sequential multi-scale perception (SMSP) algorithm, which enables WiPe-GAI to construct a feature channel state information (CSI) matrix. This is then fed into a trained neural network to predict the user's skeleton, to accurately capture the user's posture in the physical space. By integrating the user's prompts with the predicted skeleton, WiPe-GAI then guides the GAI model to generate the corresponding virtual characters for the user. Compared to image-guided AIGC, WiPe-GAI not only enhances privacy by reducing the exposure of users under the camera, but also expands service coverage through the ubiquitous availability of wireless signals. Furthermore, considering the limited resources of the VSP deployed in the mobile edge network, we design an incentive mechanism based on pricing for this framework and propose a diffusion model based approach to generate the optimal pricing strategy. This strategy maximizes the user's utility, while encouraging the VSP to actively participate in service provision, thereby ensuring the efficient operation of WiPe-GAI. In summary, the main contributions of this paper are as follows.

\begin{itemize}
\item We design a unified framework deployed in resource-constrained mobile edge networks, which combines wireless perception and GAI to provide AIGC to users. It also incorporates an incentive mechanism to ensure its economical operation.

\item In the developed framework, we propose a novel SMSP algorithm, which sequentially performs large-scale and small-scale perception on the user to construct a CSI feature matrix for user skeleton prediction. During this process, the perception at different scales cooperates by sharing perception results, thus enhancing the overall perception performance.

\item To ensure the economical operation of the framework, we design an incentive mechanism based on pricing and propose a diffusion model-based approach to generate an optimal pricing strategy. Through this strategy, users can maximize their utility while the VSP with limited resources is encouraged to participate actively in AIGC provisioning.

\item The experimental results validate the effectiveness of the proposed framework. That is, WiPe-GAI can accurately predict the user's skeleton and generate the corresponding virtual character for the user. Moreover, the proposed diffusion-based method can effectively generate the optimal strategy that not only yields greater user utility than existing methods, but also ensures VSP's participation, thereby enhancing the efficiency of the framework.

\end{itemize}

This paper is structured as follows. Section II reviews related works. Section III presents the overall framework and details the design of the framework. The evaluation is given in Section IV. Section V summarizes the paper.

\section{Related work}
In this section, we provide a brief review of the related works about wireless perception, diffusion model, and pricing-based incentive mechanisms.

\subsection{Wireless Perception}

Wireless perception involves using various signal processing techniques to extract features from wireless signals. These features are then analyzed to achieve human perception, including localization~\cite{wang2016csi}, behavior and gesture recognition~\cite{tan2020enabling}, and even imaging~\cite{karanam20173d}. In~\cite{zhao2018rf}, the authors proposed a novel convolution neural network (CNN) architecture to condense the spatial-temporal information in wireless signals, enabling the conversion of frequency modulated continuous wave (FMCW) signals to human skeleton. This approach has also been extended to through-wall scenarios~\cite{zhao2018through}. The radio-frequency identification (RFID) can also be used for human skeleton estimation. For instance, in~\cite{yang2020rfid}, the authors first calibrated the phase of RFID data and imputed the missing data via tensor completion. On this basis, they estimated the spatial rotation angle of each human limb and utilize the angles to reconstruct human pose. In addition to RFID, some other researchers have also proposed methods for converting WiFi signals to user skeleton~\cite{guo2019signal}. For example, in ~\cite{zhou2023metafi++}, the authors designed a shared convolutional module and a transformer, which explores the spatial information of human pose via self-attention and maps the WiFi CSI to human skeleton. The authors in~\cite{wang2019person} developed a deep learning approach, which takes the obtained WiFi signals as input and utilizes annotations on two-dimensional images to achieve human body segmentation and pose estimation in an end-to-end manner. Unlike these WiFi based methods, which lacks targeted processing of wireless signals, we introduce the SMSP algorithm in this paper to fully exploit the information contained in CSI, thereby predicting more accurate human skeleton.
\vspace{-0.3cm}
\subsection{Diffusion Model}
The diffusion model is a type of deep generative model~\cite{du2023beyond}, which can generate the sample by gradually learning the reverse diffusion process~\cite{croitoru2023diffusion}. This model is widely used in image generation~\cite{ho2022cascaded}. For instance, the authors in~\cite{ma2023unified} proposed a unified multi-modal latent diffusion model, which takes texts and images containing specified subjects as the input and generates customized images with the subjects. By introducing cross-attention layers, the authors in~\cite{rombach2022high} transformed the diffusion model into a generator for general conditional inputs, making it possible for high-resolution synthesis in a convolutional manner. Additionally, the authors in~\cite{cheng2023layoutdiffuse} achieved high perceptual quality image generation with less data, by adopting a novel neural adapter based on layout attention and task-aware prompts. Besides image generation, recent works apply the diffusion model to behavior cloning, policy regularization~\cite{wang2022diffusion}, and optimization problem solving~\cite{du2023ai}. Unlike existing works that focus on generating images, we propose to use the diffusion model to generate optimal pricing strategy for users and VSPs, thereby ensuring the efficient operation of WiPe-GAI.
\begin{figure*}[t]
	\centering
	\includegraphics[width=0.9\textwidth]{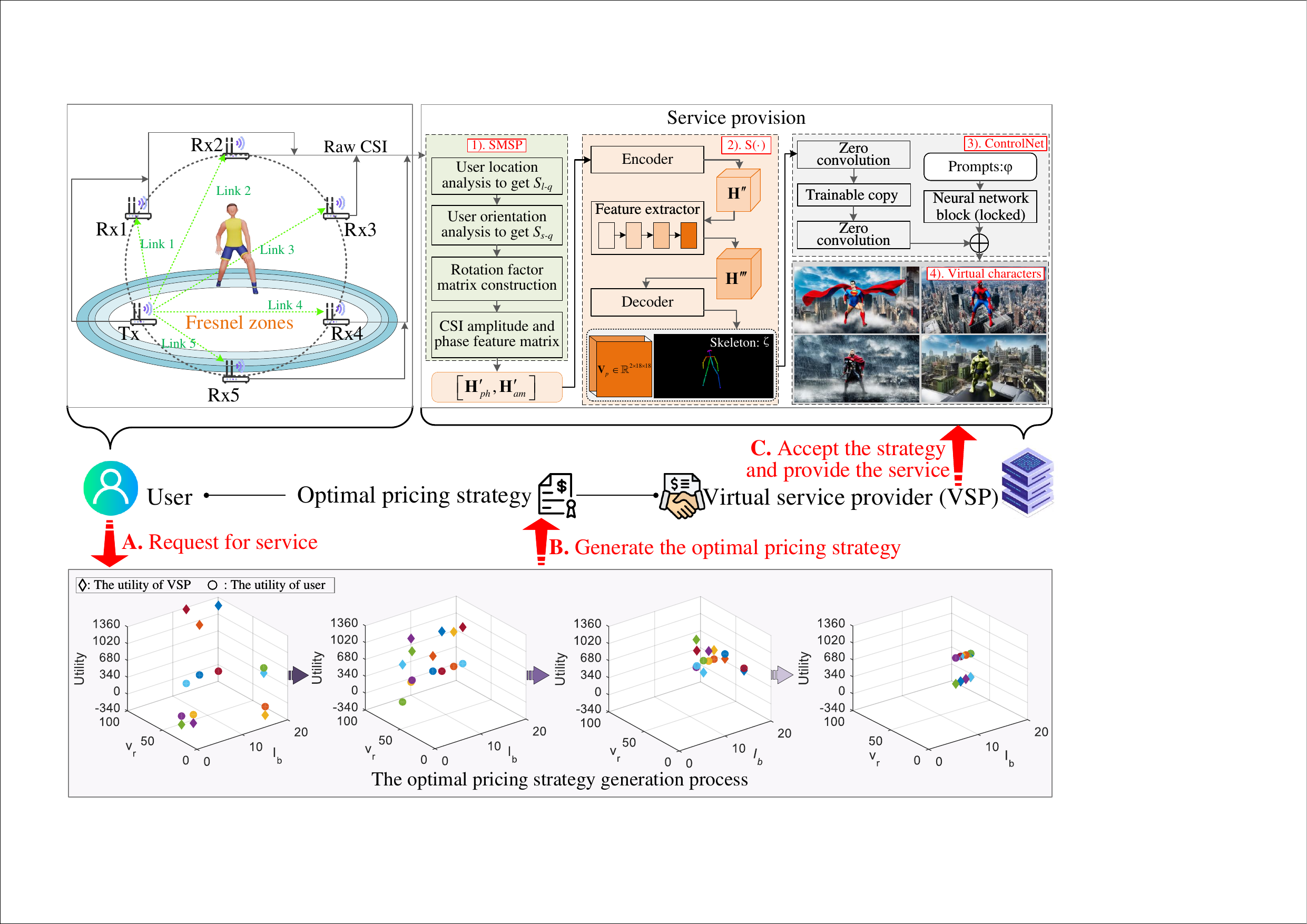}%
	\caption{The structure of the WiPe-GAI framework. When the user initiates the service request, WiPe-GAI employs the proposed diffusion model-based method to generate the pricing strategy, as four figures at the bottom show. It can be observed that, through the iterations, the pricing gradually converges to the optimal position, thereby forming the final pricing strategy. If the utility brought by this pricing strategy meets the requirements of the VSP, the VSP subsequently provides AIGC to the user.}
	\label{STCT}
\end{figure*}
\subsection{Pricing-based Incentives}
In wireless network ecosystems, pricing strategies are often used in building incentive mechanisms, with the aim of enhancing the utility of the strategy provider~\cite{luong2018applications}. For instance, the authors in~\cite{zhao2020intelligent} developed a stochastic game to simulate the dynamics between users and the access point (AP). Here, the AP establishes a price to maximize its utility, while users strategize their offloading to minimize both latency and costs. Moreover, authors in~\cite{qian2020multi} employed the Stackelberg pricing game to facilitate spectrum trading between mobile network operators (MNOs) and wireless spectrum providers (WSPs), aiming to simultaneously maximize the payoffs for both MNOs and WSPs. Their results confirmed the achievement of this goal and that the Stackelberg equilibrium can be reached. In another study~\cite{yang2022joint}, the authors introduced a pricing strategy to stimulate content caching among device-to-device (D2D) users and proposed four algorithms based on varied pricing and cache reward methods. The results indicate that a uniform pricing scheme with linear rewards is ideal for high cache quality scenarios, while the discriminatory pricing scheme with nonlinear rewards better serves scenarios demanding more evenly distributed cache content. Additionally, focusing on vehicular ad hoc networks (VANETs), the authors in~\cite{qian2020leveraging} presented a pricing strategy by considering both cellular base station's revenue and network throughput. Through extensive tests, they showed that the proposed algorithm can improve the total transmission rate of VANET by at least 20\% compared with the random selection approach. Inspired by these works, this paper aims to incentivize the VSP to actively participate in service provision, by designing an effective pricing strategy for users and VSPs.
\begin{table*}\small
\centering
\caption{{Some key notations}}
\begin{tabular}{c |c c |c c}
\hline
\textbf{Section} & \textbf{Notation} & \textbf{Definition} & \textbf{Notation} & \textbf{Definition} \\
\hline
\multirow{6}*{Perception}&$M$ &Total number of antennas  &$N$  &Total number of subcarriers\\  
\cline{2-5}  
&$k$ &Antenna spacing  &$c$  &Signal propagation speed\\
\cline{2-5}  
&$U$ &Total number of measurements  &$L$  &Total number of propagation paths\\
\cline{2-5}
&$\theta$ &Signal AoA  &$\tau$  &Signal ToF\\
\cline{2-5}
&$Q$ &Total number of receivers  &${\bf{I}}$  &Identity matrix\\
\cline{2-5}
&${{\bf{F}}_q}$ &Rotation matrix  &$\left[ {{{{\bf{{ H}'}}}_{ph}},{{{\bf{{ H}'}}}_{am}}} \right]$  &CSI feature matrix\\

\hline
\multirow{3}*{Skeleton}&${\bf{B}}\left(  \cdot  \right)$ &Neural network for converting ${\bf{V}}$ into ${\bf{V'}}$  &${\bf{S}}\left(  \cdot  \right)$  &Neural network for predicting ${\bf{V_{p}}}$\\  
\cline{2-5}
&${\bf{V}}$ &Video data  &${\bf{V'}}$  &Pose adjacent matrix\\
\cline{2-5}
\multirow{2}*{extraction}&${\bf{{H}''}}$ &Output of encoder &${\bf{{H}'''}}$  &Output of feature extractor\\
\cline{2-5}
&${\bf{V_{p}}}$ &Predicted skeleton &${{\cal L}}{}_{MSE}$  &Loss function\\

\hline
\multirow{3}*{Incentive}&$v_{r}$ &Price for per unit of QoS paid by user  &$Q_{t}$  &QoS  \\  
\cline{2-5}
&${\bf{I_{b}}}$ &Base fee provided by user to VSP &${\bf{v_{c}}}$  &Unit cost of computing resources\\
\cline{2-5}
\multirow{2}*{mechanism}&${\bf{v_{m}}}$ &User's gain per unit of QoS  &${\bf{E_{t}}}$  &Total computing resources\\
\cline{2-5}
&${\bf{U_{th}}}$ &Utility threshold of VSP &$T$  &Number of rounds to add noise\\
\hline
\end{tabular}
\end{table*}

\section{System Design}
In this section, we first provide an overview of the proposed WiPe-GAI. Subsequently, we introduce the key components, including SMSP algorithm, the user skeleton extraction, and the GAI based virtual character generation. Finally, we present the design of the pricing-based incentive mechanisms and the generation of the optimal pricing strategy based on the diffusion model.
\subsection{System Overview}
By taking virtual interactive gaming as an example, Fig.~\ref{STCT} illustrates the proposed framework, which involves two parties, i.e., the users (service requester) and the VSP deployed in mobile edge networks, as well as three core steps, represented by A, B, and C, respectively. Specifically, when the user initiates a service request, the WiPe-GAI first employs the proposed diffusion model based approach to generate the optimal pricing strategy and presents it to the VSP. Once the utility brought by the strategy meets the requirements of the VSP, the VSP provides AIGC services to the user. As the service provision part in Fig.~\ref{STCT} shows, the VSP first runs the proposed SMSP algorithm to construct the CSI feature matrix, which, unlike the raw CSI data, emphasizes the information relevant to the user. Subsequently, leveraging the trained neural network (denoted as ${\bf{B}}\left(  \cdot  \right)$), the extracted CSI feature matrix is converted into a skeleton, accurately representing the user's posture in the physical world. Lastly, the VSP uses the acquired skeleton to guide GAI to generate a corresponding virtual character for the user. In contrast to other guiding strategies based on images or videos, WiPe-GAI not only offers better protection of user's privacy but also has a wider coverage due to the ubiquity of wireless signals~\cite{yang2013rssi}. Meanwhile, the pricing-based incentive mechanism and the corresponding generated optimal pricing strategy ensure the entire framework operates efficiently. Next, we will detail the designs of the proposed WiPe-GAI. To facilitate the description, we summarize the main notation in Table I.

\subsection{Sequential Multi-scale Perception}
\subsubsection{Large-scale Perception}
Upon receiving the service request from the user, the VSP employs the wireless nodes around the user to perform SMSP by transmitting and receiving wireless signals. Using the captured wireless signals, the first step is to perform large-scale perception. Concretely, assuming that the device located at $\left[ {{x_t},{y_t}} \right]$ transmits the signals modulated by the orthogonal frequency division multiplexing (OFDM) technique, while the wireless node located at $\left[ {{x_q},{y_q}} \right]$ utilizes a uniform linear antenna array to receive signals. Then, the CSI obtained by the $q$-th receiver can be expressed as
\begin{align}\label{eq1}
{\bf{H}} = \left[ \begin{array}{l}
H{}_{1,1}{\ } \cdots {\ }H{}_{1,N}\\
{\quad } \vdots {\quad} \ddots {\quad} \vdots \\
H{}_{M,1} \cdots {\rm{ }}H{}_{M,N}
\end{array} \right],
\end{align}
where $H{}_{m,n}$ is the CSI extracted from the $m$-th antenna and the $n$-th subcarrier, $M$ represents the number of antennas at the receiver, and $N$ represents the number of subcarriers. Each element in matrix ${\bf{H}}$ is the sum of the CSI of all the signal propagation paths~\cite{yang2013rssi}. For any given specified propagation path $l$, the corresponding CSI can be written as 
\begin{align}\label{eq2}
H_{m,n}^{\left[ l \right]} = \alpha _{m,n}^{\left[ l \right]}{e^{ - j2\pi {f_n}\left[ {\tau _q^{\left[ {l1} \right]} + {{\left( {m - 1} \right)k\sin \left( {\theta _q^{\left[ l \right]}} \right)} \mathord{\left/
 {\vphantom {{\left( {m - 1} \right)k\sin \left( {\theta _q^{\left[ l \right]}} \right)} c}} \right.
 \kern-\nulldelimiterspace} c}} \right]}}{e^{ - j\varepsilon }}{\rm{ + }}\eta _{m,n}^{\left[ l \right]},
\end{align}
where $\alpha _{m,n}^{\left[ l \right]}$ represents the attenuation introduced by the propagation path, ${f_n}$ is the frequency of the $n$-th subcarrier, $\tau _q^{\left[ {l1} \right]}$ is the time of flight (ToF) of the signal arriving at the reference antenna, $k$ represents the antenna spacing at the receiver (assumed to be half-wavelength~\cite{kotaru2015spotfi}), $\theta _q^{\left[ l \right]}$ represents the signal angle of arrival (AoA), $c$ is the signal propagation speed, ${e^{ - j\varepsilon }}$ represents the phase error, and $\eta _{m,n}^{\left[ l \right]}$ is the noise. 

As it can be observed from (\ref{eq2}), for each propagation path, the signal AoA is encoded in the phase difference between the antennas, while the ToF is embedded in the phase difference between the subcarriers. Consequently, based on ${\bf{H}}$, the two-dimensional multiple signal classification algorithm is used here to jointly estimate the AoA and ToF of the propagation path. The basic idea of this algorithm is the eigenstructure analysis of a correlation matrix ${{\bf{R}}_{\rm{X}}}$, which is giving
\begin{align}\label{eq3}
{{\bf{R}}_{\rm{X}}}{\rm{ = E}}\left[ {{\bf{X}}{{\bf{X}}^\dag }} \right] = {\bf{A}}{{\bf{R}}_S}{{\bf{A}}^\dag } + {\sigma ^2}{\bf{I}},
\end{align}
where ${\bf{X}}\in {\mathbb{R}^{M' \times N'}}$ is obtained by conducting spatial smoothing on the ${\bf{H}}$, the superscript $\dag $ is the conjugate transpose operator, ${\bf{A}}$ is the array manifold corresponding to ${\bf{X}}$, ${{\bf{R}}_S}$ is the correlation matrix of the signal matrix, ${\bf{I}}$ is the identity matrix, and ${\sigma ^2}$ is the variance of noise. The matrix ${{\bf{R}}_{\rm{X}}}$ has 
$M'$ eigenvalues, among which the larger ones correspond to eigenvectors that form the signal subspace ${{{\bf{E}}_S}}$. According to the information theoretic criteria~\cite{wax1985detection}, the number of large eigenvalues, denoted as $L$, can be estimated by minimizing
\begin{align}\label{eq4}
MDL\left( L \right) &=  - \log {\left[ {\frac{{\prod\limits_{i = L + 1}^{M'} {\lambda _i^{{1 \mathord{\left/
 {\vphantom {1 {\left( {M' - L} \right)}}} \right.
 \kern-\nulldelimiterspace} {\left( {M' - L} \right)}}}} }}{{\frac{1}{{M' - L}}\sum\limits_{i = L + 1}^{M'} {{\lambda _i}} }}} \right]^{\left( {M' - L} \right)U}}\\ \notag
 &+ \frac{1}{2}L\left( {2M' - L} \right)\log \left( U \right),
\end{align}
where ${\lambda _i}$ is the $i$-th largest eigenvalue, and $U$ is the number of observations\footnote{This value is determined based on the data transmission rate of the wireless nodes. For instance, assuming the node is a commonly used WiFi device with a data packet transmission frequency of 400 Hz. Then, based on the channel coherence time~\cite{vasisht2016decimeter}, $U$ can be set as $400 \times 0.84 \approx 34$.}. Apart from the signal subspace, the remaining eigenvectors form the noise subspace ${{{\bf{E}}_N}}$, which is orthogonal to the steering matrix ${{\bf{a}}^{\dag}}\left( {\theta ,\tau } \right)$ extracted from ${\bf{X}}$. Using this orthogonality, we have 
\begin{align}\label{eq5}
{P_s}\left( {\theta ,\tau } \right) = \frac{1}{{{{\bf{a}}^{\dag}}\left( {\theta ,\tau } \right){{\bf{E}}_N}{\bf{E}}_N^{\dag}{\bf{a}}\left( {\theta ,\tau } \right)}},
\end{align}
through which the joint AoA and ToF estimation for each signal propagation path is achieved by traversing AoA and ToF, i.e., ${\theta}$ and ${\tau}$, respectively. In this way, the VSP uses the CSI obtained from each receiver to estimate the AoA and ToF corresponding to the user induced reflection. By combining these estimated parameters, along with the locations of the transceivers, the VSP calculates the user's physical location, and here we denote it as $\left[ {{x_{us}},{y_{us}}} \right]$. 

According to the Fresnel Zone Theory~\cite{zhang2021fresnel}, the user's posture has a greater influence on nearby wireless links. This implies that the CSI obtained from wireless links closer to the user carries more detailed information regarding the user's posture. Hence, the VSP calculates the distance between the user and each wireless link. By using the link formed by the $q$-th receiver and transmitter as an example, this distance is
\begin{align}\label{eq6}
{D_{_q}} = \frac{{\left| {\Upsilon {x_{us}} + \Upsilon '{y_{us}} + \left( {{x_q} - {x_t}} \right){y_t} - \left( {{y_{us}} - {y_t}} \right){x_t}} \right|}}{{\sqrt {{\Upsilon ^2} + {{\Upsilon '}^2}} }},
\end{align}
where $\Upsilon  = {y_q} - {y_t}$, and $\Upsilon ' = {x_t} - {y_q}$. On this basis, the score of large-scale perception is calculated according to the computed distance as follows:
\begin{align}\label{eq7}
{S1_{q}} = {{\min \left\{ {{D_q}} \right\}} \mathord{\left/
 {\vphantom {{\min \left\{ {{D_q}} \right\}} {{D_q}}}} \right.
 \kern-\nulldelimiterspace} {{D_q}}},
\end{align}
where ${\min \left\{ {{D_q}} \right\}}$ represents the minimum value among $Q$ distances, $q = 1,\ldots , Q$, and $Q$ is the total number of receivers which are involved in the perception. As shown in (\ref{eq7}), links closer to the user yield higher scores due to the richer information that they contain. These scores will later be utilized as weights during the construction of the CSI feature matrix, and hence ensuring that links with more information play a more pivotal role in skeleton prediction. In this manner, the VSP accomplishes large-scale perception of the user, and the obtained user's location will then be used to assist in the subsequent small-scale perception.
\begin{figure*}[t]
	\centering
	\includegraphics[width=1\textwidth]{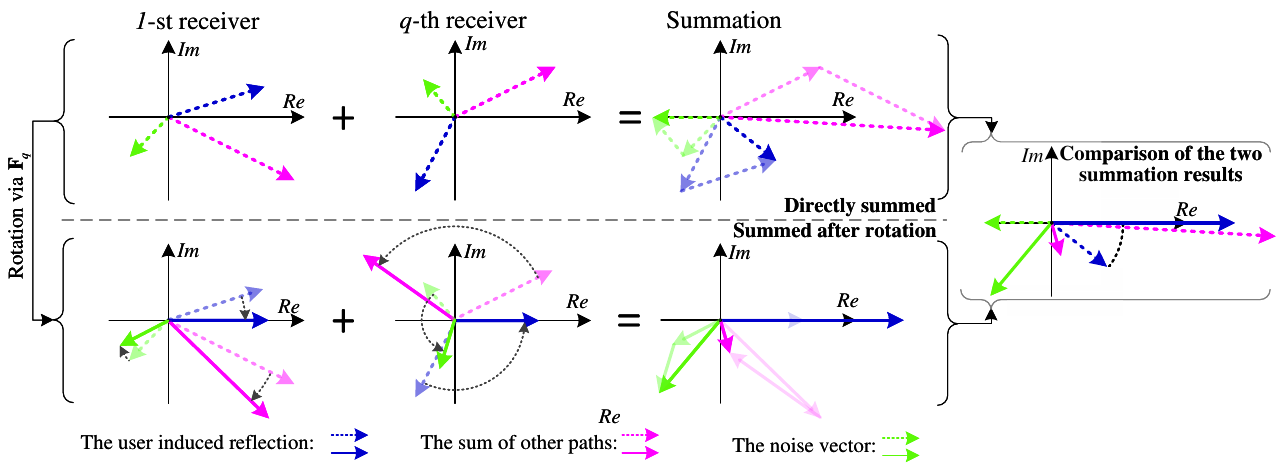}%
	\caption{The enhancement process of the CSI induced by the user before the feature matrix construction. As illustrated in the first row of the figure, if the raw CSIs are directly summed, the inconsistent phases of CSIs obtained from different receivers may reduce the proportion of user-induced reflections in the summed CSI, consequently degrading the perception performance. In WiPe-GAI, we propose to rotate the CSIs corresponding to the user to the same direction before summation, as depicted in the second row of the figure, thereby circumventing such an issue.}
	\label{CSIStrenth}
\end{figure*}
\subsubsection{Small-scale Perception} To improve the accuracy of the extracted user skeleton, the VSP further conducts small-scale perception of users to obtain the CSI that contains more detailed information about user's posture. Inspired by the Fresnel Zone Theory and the impact of user's orientation and behavior on the wireless link, the VSP analyzes the signal fluctuation characteristic with the help of large-scale perception result, to achieve small-scale perception. 

Specifically, the VSP first utilizes $\left[ {{x_{us}},{y_{us}}} \right]$ and $\left[ {{x_q},{y_q}} \right]$ to calculate the direction of the user relative to the $q$-th receiver, denoted as ${\theta '_q}$. Then, the VSP uses ${\theta '_q}$ to construct a phase weight for the CSI of the $m$-th antenna and the $n$-th subcarrier, which is
\begin{align}\label{eq8}
{\rm{ }}{w_{m,n}}\left( {\theta '_q} \right) = {e^{j2\pi {f_n}\frac{{\left( {m - 1} \right)k\sin \left( {\theta '_q} \right)}}{c}}}.
\end{align}
By using this weight, the power received along a beam in the ${\theta '_q}$ direction of $q$-th receiver at time $u$ can be calculated as
\begin{align}\label{eq9}
{P_w^{\left[ u \right]}}\left( {\theta '_q} \right) = {\left| {\sum\limits_{m = 1}^M {\sum\limits_{n = 1}^N {{w_{m,n}} \cdot {H_{m,n}}} } } \right|^2}.
\end{align}
Assuming the power in (\ref{eq9}) is obtained at time $u$, and then for a power stream containing $U$ observations, the VSP employs unbiased variance to characterize the fluctuation features of the wireless link during this period of time as follows:
\begin{align}\label{eq10}
S_{\theta '_q}^2 = \frac{1}{{U - 1}}\sum\limits_{u = 1}^U {{{\left[ {P_w^{\left[ u \right]}\left( {\theta '_q} \right) - {{\bar P}_w}\left( {\theta '_q} \right)} \right]}^2}}, 
\end{align}
where ${{{\bar P}_w}\left({\theta '_q} \right)}$ is the average power value during this period. By doing so, the score of small-scale perception is obtained by calculating the variance of each wireless transmission link as follows:
\begin{align}\label{eq11}
{S2_{q}} = {{S_{\theta '_q}^2} \mathord{\left/
 {\vphantom {{S_{\theta _q^{\left[ l \right]}}^2} {\max \left\{ {S_{\theta '_q}^2} \right\}}}} \right.
 \kern-\nulldelimiterspace} {\max \left\{ {S_{\theta '_q}^2} \right\}}},
\end{align}
where $\max \left\{ {S_{\theta '_q}^2} \right\}$ is the maximum among $Q$ variances. From (\ref{eq8}) to (\ref{eq11}), it can be seen that a link influenced more significantly by the user's posture (i.e., with higher link fluctuations) tends to contain more information~\cite{zhang2021fresnel}, subsequently yielding a higher score. With the help of large-scale perception results, the VSP finishes the small-scale perception of the user and obtains the corresponding score, which will be combined with the large-scale score later to create the CSI feature matrix used for skeleton generation.

To further improve the user skeleton extraction performance by combining CSI from all receivers, the VSP performs more processing on the original CSI data to enhance the user induced reflection before constructing the CSI feature matrix. We use the case with two receivers as an example. For the $m$-th antenna and the $n$-th subcarrier, the CSI obtained by the $1$-st and the $q$-th receivers are shown in Fig.~\ref{CSIStrenth}, where the blue arrowed line represents the CSI of the user induced reflection signal, the red arrowed line is the CSI corresponding to the sum of the signals from all other paths, and the green arrowed line is the noise. As shown by the first row in Fig.~\ref{CSIStrenth}, if directly summed, the phase inconsistency of the CSI among different receivers may weaken the user induced reflection, thereby reducing the perception performance. To circumvent such an issue, the VSP needs to rotate the CSI of the user obtained by each receiver to the same direction. Recall that the phase (i.e., the angle between the blue vector and the $Re$-axis) of the CSI corresponding to the user induced reflection is determined by the ToF and initial phase, while the amplitude (i.e., the magnitude of the blue vector) is determined by the reflection coefficient~\cite{zeng2021exploring}. Therefore, using the estimated AoA and ToF of the user induced reflection, the VSP builds a rotation factor matrix. For the $q$-th receiver, the matrix is
\begin{align}\label{eq12}
{{\bf{F}}_q} = \left[ {\begin{array}{*{20}{c}}
{{\rm{ }}F_q^{\left[ {1,1} \right]}}& \cdots &{F_q^{\left[ {1,N} \right]}}\\
 \vdots & \ddots & \vdots \\
{F_q^{\left[ {M,1} \right]}}& \cdots &{F_q^{\left[ {M,N} \right]}}
\end{array}} \right],
\end{align}
where 
\begin{align}\label{eq13}
F_q^{\left[ {m,n} \right]} = {e^{j2\pi {f_n}\left[ {\tau _q^{\left[ {l1} \right]} + \frac{{\left( {m - 1} \right)k\sin \left( {\theta _q^{\left[ l \right]}} \right)}}{c}} \right]}}.
\end{align}
Next, ${\bf{H}}$ is multiplied with ${{\bf{F}}_q}$ to rotate the CSI induced by the user to the positive direction of the $Re$-axis, as
\begin{align}\label{eq14}
{{\bf{H'}}_q} = {\bf{H}} \circ {{\bf{F}}_q},
\end{align}
where $\circ $ is the Hadamard product operator. By performing this operation to all receivers, the CSI corresponding to the user induced reflection received by each receiver will be rotated towards the same direction. 
\begin{figure*}[t]
	\centering
	\includegraphics[width=1\textwidth]{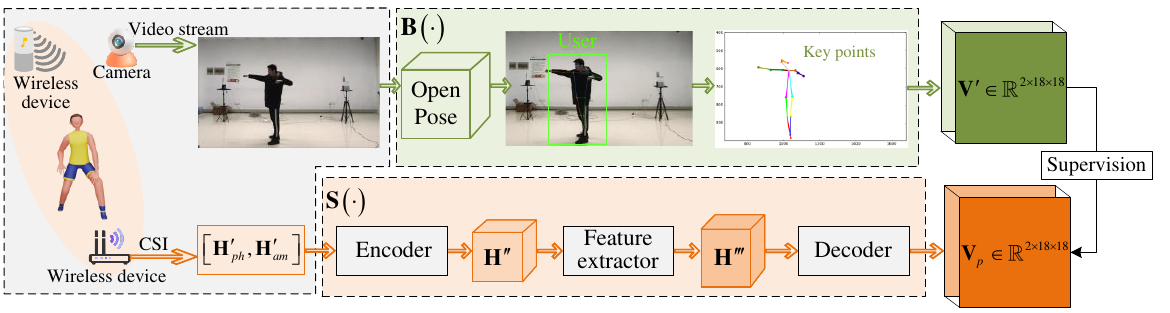}%
	\caption{The training process of the network which converts CSI feature matrix to user skeleton. Note that during the training process, the VSP needs to use a camera to acquire ${\bf{V'}}$, which serves as supervision to optimize the neural network ${\bf{S}}\left(  \cdot  \right)$. However, during the operation of framework WiPe-GAI, the VSP only needs the wireless signals to generate the user's skeleton, without the assistance of a camera. This makes the proposed framework applicable to more scenarios where cameras are not suitable. }
	\label{CSI2SKLT}
\end{figure*}
As demonstrated in the second row of Fig.~\ref{CSIStrenth}, this ensures that no attenuation occurs during the summation process. Subsequently, the rotated CSI from each receiver is weighted by the scores acquired by the SMSP algorithm, which are then aggregated to construct the CSI amplitude and phase feature matrix, respectively denoted as:
\begin{align}\label{eq15}
\left\{ \begin{array}{l}
{{{\bf{{H}'}}}_{ph}} = \sum\limits_{q = 1}^Q {\left( {S{1_q} + S{2_q}} \right)angle\left( {{{{\bf{H'}}}_q}} \right)} \\
{{{\bf{{H}'}}}_{am}} = \sum\limits_{q = 1}^Q {\left( {S{1_q} + S{2_q}} \right)abs\left( {{{{\bf{H'}}}_q}} \right)},
\end{array} \right.
\end{align}
where $angle\left\{  \cdot  \right\}$ and $abs\left\{  \cdot  \right\}$ are phase and 
amplitude extractor, respectively. From (\ref{eq15}), it is clear that the derived CSI feature matrix is abundant with information about user's posture. Hence, the VSP uses these matrices to generate human skeleton data with neural networks, which will be explained in detail in the following section.

\subsection{Skeleton Extraction}
Based on the acquired CSI feature matrix, the VSP further needs to convert it into skeleton before feeding it into the GAI model for the generation of a virtual user character. To this end, the VSP utilizes a camera synchronized with the signal receiver to capture a video stream, from which the user's skeleton is extracted (via neural network ${\bf{B}}\left(  \cdot  \right)$) and used as supervision to train a neural network (denoted as ${\bf{S}}\left(  \cdot  \right)$), as shown in Fig.~\ref{CSI2SKLT}. Finally, based on the trained neural network, the VSP can convert the CSI feature matrix into user's skeleton.

Specifically, let $\left\{ {{\bf{V}},{\bf{{ H}''}}} \right\}$ be a pair of synchronized training data, where ${\bf{{H}''}}$ is composed of multiple samples of ${{\bf{{H}'}}_{ph}}$ and ${{\bf{{H}'}}_{am}}$, since the sampling rate of CSI is higher than that of the video frame. To convert the CSI data into skeleton data, the neural networks ${\bf{B}}\left(  \cdot  \right)$ and ${\bf{S}}\left(  \cdot  \right)$ are constructed. For any given data pair, ${\bf{B}}\left(  \cdot  \right)$ takes ${\bf{V}}$ as the input and outputs skeleton data containing 18 points, by using OpenPose~\cite{cao2021openpose}. After that, these 18 points are transformed into a pose adjacent matrix ${\bf{V'}}$, and we denote this process as $ \bf{B}\left( {\bf{V}} \right) \Rightarrow {\bf{V'}} \in {\mathbb{R}^{2 \times 18 \times 18}}$. At the same time, ${\bf{S}}\left(  \cdot  \right)$ takes ${\bf{{ H}''}}$ as input and predicts ${{\bf{V}}_p}$, which is denoted as $\bf{S}\left( {{\bf{{ H}''}}} \right) \Rightarrow {{\bf{V}}_p} \in {\mathbb{R}^{2 \times 18 \times 18}}$. On this basis, ${\bf{S}}\left(  \cdot  \right)$ is optimized with the supervision of ${\bf{V'}}$, to assist training. The architecture of this network is shown in Fig.~\ref{CSI2SKLT}, where ${\bf{S}}\left(  \cdot  \right)$ includes three components: encoder, feature extractor, and decoder, which are introduced below.

\begin{figure}[t]
\centering
\includegraphics[height=2.7cm]{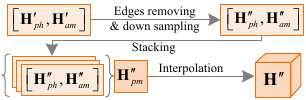} 
\caption{The structure of the encoder.} 
\label{Encoder} 
\vspace{-0.5cm}
\end{figure}

\textbf{Encoder}. This module is designed to adjust the data dimension of ${\bf{{H}''}}$ through operations such as data deletion and interpolation. In this paper, the CSI is collected using an IEEE 802.11ac wireless node, with one antenna at the transmitter and four antennas at the receiver. One of the receiver's antennas is used for phase calibration and the remaining ones for ${\bf{{ H}''}}$ construction. Because of the different sampling rates between the camera and receiver, one image is used to match three CSI measurements. Therefore, we have ${{\bf{{ H}'}}_{ph}} \in {\mathbb{R}^{256 \times 3}}$ and ${{\bf{{ H}'}}_{am}} \in {\mathbb{R}^{256 \times 3}}$, where 256 represents the number of subcarriers and 3 represents the number of antennas. Subsequently, the encoder removes the CSI corresponding to subcarriers at the bandwidth edges and performs down-sampling to convert $\left[ {{{{\bf{{ H}'}}}_{ph}},{{{\bf{{ H}'}}}_{am}}} \right] \in {\mathbb{R}^{512 \times 3}}$ to $\left[ {{{{\bf{{ H}''}}}_{ph}},{{{\bf{{ H}''}}}_{am}}} \right] \in {\mathbb{R}^{150 \times 3}}$. On this basis, three $\left[ {{{{\bf{{ H}''}}}_{ph}},{{{\bf{{ H}''}}}_{am}}} \right]$ are directly stacked to obtain ${{\bf{{ H}''}}_{pm}} \in {\mathbb{R}^{150 \times 3 \times 3}}$. After that, ${{\bf{{ H}''}}_{pm}} \in {\mathbb{R}^{150 \times 3 \times 3}}$ is interpolated to obtain ${\bf{{H}''}} \in {\mathbb{R}^{150 \times 144 \times 144}}$. Specifically, assuming that the values of four adjacent elements in ${{\bf{{ H}''}}_{pm}}$ are ${h''_{11}}$, ${h''_{12}}$, ${h''_{21}}$, and ${h''_{2}}$, respectively, and their corresponding coordinates are $\left[ { \cdot ,{r_1},{c_1}} \right]$, $\left[ { \cdot ,{r_1},{c_2}} \right]$, $\left[ { \cdot ,{r_2},{c_1}} \right]$, and $\left[ { \cdot ,{r_2},{c_2}} \right]$, respectively. Using these four elements, the element obtained through interpolation located at $\left[ { \cdot ,{r},{c}} \right]$ is
\begin{align}\label{eq16}
{h'''_{rc}} &= \left[ {{h''_{11}}\left( {{r_2} - r} \right)\left( {{c_2} - c} \right) + {h''_{21}}\left( {r - {r_1}} \right)\left( {{c_2} - c} \right)} \right]\notag \\ 
&+ \left[ {{h''_{12}}\left( {{r_2} - r} \right)\left( {c - {c_1}} \right) + {h''_{22}}\left( {r - {r_1}} \right)\left( {c - {c_1}} \right)} \right],
\end{align}
where ${r_1} < r < {r_2}$ and ${c_1} < c < {c_2}$. At last, ${\bf{{H}''}}$ is fed into the next module for feature extraction. The structure of the encoder is shown in Fig.~\ref{Encoder}.

\begin{figure*}[t]
	\centering
	\includegraphics[width=0.85\textwidth]{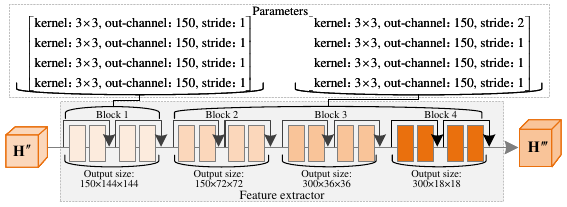}%
	\caption{The structure and parameters of the feature extractor.}
	\label{Fextr}
\end{figure*}

\textbf{Feature extractor}. Based on ${\bf{{H}''}}$, a feature extractor is used to learn the effective features for user posture estimation. As deeper networks are known to have greater feature learning capabilities, the VSP could use them to fully unleash the feature information contained within ${\bf{{H}''}}$. However, the potential risk associated with deeper networks, i.e., the gradient vanishing or exploding in deep convolutional layers caused by the chain rule in the back-propagation optimization, must also be taken into consideration. The ResNet~\cite{he2016deep}, a widely-used network in deep learning, can alleviate this problem through the use of shortcut connections and residual blocks. Hence, the VSP stacks four ResNets basic blocks to form the feature extractor, as shown in Fig.~\ref{Fextr}, for learning features ${\bf{{H}'''}} \in {\mathbb{R}^{300 \times 18 \times 18}}$. Note that each convolutional layer is followed in succession by a batch normalization layer~\cite{ioffe2015batch} and a rectified linear unit activation layer~\cite{krizhevsky2017imagenet}. 

\begin{figure}[t]
\centering
\includegraphics[height=2cm]{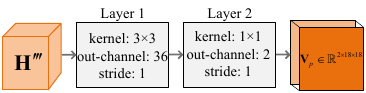} 
\caption{The structure of the decoder.} 
\label{Decoder} 
\end{figure}

\textbf{Decoder}. The purpose of the decoder is to perform shape adaptation between ${\bf{{H}'''}}$ and ${\bf{V'}}$. As explained for the encoder, ${\bf{V'}}$ is a tensor of size $2 \times 18 \times 18$, and the decoder takes ${\bf{{H}'''}}$ as the input and predicts matrix ${{\bf{V}}_p}$ which has the same size as ${\bf{V'}}$. To accomplish this, the decoder utilizes two convolutional layers, as depicted in Fig.~\ref{Decoder}, where the first layer primarily extracts the channel-wise information, and the second layer reorganizes the spatial information of ${\bf{{H}'''}}$ using $1 \times 1$ convolutional kernels. During the training phase, ${\bf{B}}\left( {\bf{V}} \right) \Rightarrow {\bf{V'}}$ is used as the supervision and ${\bf{S}}\left( {{\bf{{ H}''}}} \right) \Rightarrow {{\bf{V}}_p}$ is the prediction. Hence, the loss function is set as the mean squared error (MSE) between ${\bf{V'}}$ and ${{\bf{V}}_p}$, which is:
\begin{align}\label{eq17}
{{\cal L}}{}_{MSE} = \left\| {{{\bf{V}}_p} - {\bf{V'}}} \right\|_2^2.
\end{align}

Under the above configurations, the network is trained for 20 epochs using the Adam optimizer with an initial learning rate of 0.001 and a batch size of 32. The learning rate is decayed by 0.5 at the 5-th, 10-th, and 15-th epochs. Once the training is finished, the model shall be able to predict ${{\bf{V}}_p}$ using only the CSI feature matrix. Finally, the diagonal elements from ${{\bf{V}}_p}$ are extracted and paired to get the predicted user skeleton. The pairing process can be denoted as 
\begin{align}
\left\{ \begin{array}{l}\label{eq18}
{X_p} = {{\bf{V}}_p}_{\left( {1,p,p} \right)},{\rm{ }}p \in \left[ {1,18} \right] \\
{Y_p} = {{\bf{V}}_p}_{\left( {2,p,p} \right)},{\rm{ }}p \in \left[ {1,18} \right],
\end{array} \right.
\end{align}
where ${X_p}$ and ${Y_p}$ are the coordinates of the predicted skeleton. 
\subsection{Generative AI Based Content Generation}
After obtaining user skeleton, the VSP needs to further generate virtual characters or even specific background based on the user's requests. So far, many GAI models have been proposed for such tasks. In this paper, the VSP is deployed at the network edge to provide such services to users. Considering the size of the training dataset, training time, and deployability, ControlNet~\cite{zhang2023adding} is used to generate virtual characters for users. However, unlike existing work that employs image as guidance~\cite{li2023blip}, WiPe-GAI utilizes the predicted user skeleton to guide ControlNet to generate the virtual character and the corresponding background for the user.

\begin{figure}[t]
\centering
\includegraphics[height=4.5cm]{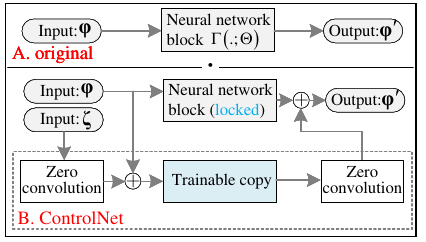} 
\caption{A comparison between the original network and ControlNet. Building upon the original network, ControlNet creats a trainable block for external condition training. Meanwhile, the neural network is connected to the zero convolution layers, where both weight and bias are initialized with zeros, and subsequently transform from zero to the optimal parameters through the training.} 
\label{CTNT} 
\vspace{-0.5cm}
\end{figure}

Specifically, assuming a feature matrix is ${\bm{\varphi }}$, a neural network $\Gamma \left( {\cdot;{\bf{\Theta}} } \right)$, where ${\bf{\Theta}}$ is a set of network parameters, can transform the feature matrix ${\bm{\varphi }}$ into another feature matrix ${\bm{\varphi '}}$, i.e.,  $\Gamma \left( {{\bm{\varphi }};\Theta } \right){\rm{ = }}{\bm{\varphi '}}$. This process is illustrated in part A of Fig.~\ref{CTNT}. To control neural networks in generating digital content according to the user's skeleton, VSP first locks ${\bf{\Theta}}$, and then copies and creates a trainable ${\bm{\Theta'}}$, which is trained with an external condition vector ${\bm{\zeta }}$. This operation not only mitigates the over-fitting problem due to a limited number of samples, but also maintains the quality of the content produced by the original network. After that, the neural network block is connected to a unique “zero convolution” layer, i.e., a $1\times 1$ convolution layer where both weight and bias are initialized with zeros, as shown in Fig.~\ref{CTNT}. By doing so, such a layer can gradually grow from zero to the optimal parameters through training. Therefore, the user virtual character generated based on the trained network can meet the user's needs in terms of character posture and image quality. Based on this structure, the VSP uses stable diffusion as the core neural network, with the user prompts serving as ${\bm{\varphi }}$ and the extracted user skeleton as the external condition vector ${\bm{\zeta }}$, to generate the virtual character for AIGC service provisioning. 

\subsection{Pricing-based Incentive Mechanism Design}
Given the limited resources of VSP deployed at the mobile edge networks, we propose a pricing-based incentive mechanism to ensure efficient operation of WiPe-GAI. In this mechanism, the user compensates the VSP based on the quality of both perception and virtual character generation services, to encourage the VSP's active participation. On this basis, we further propose a diffusion model based method to generate the optimal pricing strategy for the implementation of this incentive mechanism.
\subsubsection{Incentive Mechanism Design}
We design a pricing strategy to stimulate a VSP to engage actively in service provision while maximizing the benefits of users. In particular, assuming that the VSP provides perception and AIGC services to users, then the user pays a basic fee, as well as an additional fee based on the quality of service (QoS) to the VSP. Therefore, the profit of the VSP can be denoted as
\begin{align}\label{eq19}
{I_{VSP}}{{ = }}{v_r}{Q_t} + {I_b},
\end{align}
where ${v_r}$ denotes the price that the user pays to the VSP for per unit of QoS, ${Q_t}$ represents the QoS, and ${I_b}$ denotes the basic fee offered by the user to the VSP. Since the service provided by the VSP consists of wireless perception and AI-based virtual digital content generation, the QoS measure should consider the performance of both tasks. Specifically, wireless perception provides the skeleton for GAI, and then the GAI generates the virtual character with the same posture as the user in the physical world. Therefore, the following metrics are used.

\begin{itemize}
\item The reciprocal of the normalized Euclidean distance between ${{\bf{V}}_p}$ and ${\bf{V'}}$ is used as ${Q_s}$ to quantify the precision of perception. As more computing resources are allocated to perception, the VSP can engage more wireless nodes to participate in perception, leading to a more accurate skeleton. Therefore, we have ${Q_s} = {\varsigma _s}\left( {{\chi _s}} \right)$, where ${\chi _s}$ represents the computing resources allocated to the wireless perception and ${\varsigma _s}\left(  \cdot  \right)$ is the mapping relationship between computing resources and QoS. 

\item The Blind/Referenceless Image Spatial Quality Evaluator (BRISQUE) and Total Variation (TV) are utilized to assess the QoS of AIGC. Similar to wireless perception, when more computing resources are assigned to GAI, the GAI model can execute more inferences, resulting in better QoS. Hence, the QoS of AIGC is ${Q_{ag}} ={BRISQUE + TV}= {\varsigma _{brq}}\left( {{\chi _{ag}}} \right) + {\varsigma _{tv}}\left( {{\chi _{ag}}} \right)$, where ${\chi _{ag}}$ represents the resource allocation for GAI by the VSP, ${\varsigma _{brq}}\left(  \cdot  \right)$ is the mapping relationship between computing resources and BRISQUE, and ${\varsigma _{tv}}\left(  \cdot  \right)$ is the mapping relationship between computing resources and TV\footnote{These mapping relationships, including ${\varsigma _s}\left(  \cdot  \right)$, ${\varsigma _{brq}}\left(  \cdot  \right)$, and ${\varsigma _{tv}}\left(  \cdot  \right)$ are obtained by fitting real-world test results, which will be explained in detail in Section IV.}. 
\end{itemize}

Based on the above analysis, we can model the total QoS of the service as 
\begin{align}\label{eq20}
{Q_t} &= {Q_s} +  {{Q_{ag}}}= {\varsigma _s}\left( {{\chi _s}} \right) + {\varsigma _{brq}}\left( {{\chi _{ag}}} \right) + {\varsigma _{tv}}\left( {{\chi _{ag}}} \right).
\end{align}
Given ${I_{VSP}}$ and ${Q_t}$, the utility function of the VSP can be obtained as
\begin{align}\label{eq21}
{U_{vsp}} = {I_{VSP}} - \left( {{\chi _s} + {\chi _{ag}}} \right){v_c} = {v_r}{Q_t} + {I_b} - \left( {{\chi _s} + {\chi _{ag}}} \right){v_c},
\end{align}
where ${v_c}$ is the unit cost of computing resources, and ${v_r}$ is the fee paid by user for a unit QoS value. Meanwhile, for users, the utility function can be defined as 
\begin{align}\label{eq22}
{U_{us}} = {v_m}{Q_t} - \left( {{v_r}{Q_t} + {I_b}} \right) = \left( {{v_m} - {v_r}} \right){Q_t} - {I_b},
\end{align}
where ${v_m}$ is the gain per unit QoS obtained by user, which is determined by the market. Based on the aforementioned model, the pricing strategy offered by users includes $I_b$ and $v_r$, which aims to maximize user's utility and provide rational incentives for the VSP to agree to the pricing strategy. To obtain an optimal pricing strategy, we formulate an optimization problem as follows:
\begin{align}\label{eq23}
\begin{array}{l}
\mathop {\max }\limits_{{v_r},{I_b},{\chi _s},{\chi _{ag}}} {\rm{   }}{U_{us}}\left( {{v_r},{I_b},{\chi _s},{\chi _{ag}}} \right) \\ 
{\rm{s}}{\rm{.t}}{\rm{.  }}\left\{ \begin{array}{l}
{{\chi '}_s},{{\chi '}_{ag}} \in \mathop {\arg\rm {max} }\limits_{{\chi _s},{\chi _{ag}}}{\ }{U_{vsp}}\left( {{v_r},{I_b},{\chi _s},{\chi _{ag}}} \right),\\
{{\chi '}_s}{\rm{ + }}{{\chi '}_{ag}} \le {E_t},\\
{U_{vsp}}\left( {{{\chi '}_s},{{\chi '}_{ag}},{v_r},{I_b},} \right) \ge {U_{th}},
\end{array} \right.
\end{array}
\end{align}
where the first constraint is to ensure that the VSP can maximize its own utility, the second one comes from the limited computing resources of the VSP, and the third one is the utility threshold ${U_{th}}$, signifying that the VSP only participates in service provision when the expected utility exceeds this value. As demonstrated by the above model, the user maximizes their own utility through pricing, while the VSP seeks to optimize its utility by conducting resource allocation while meeting the constraints imposed by the provided pricing and limited computing resources. Therefore, the optimization problem is essentially a joint pricing and resource allocation problem. Considering the uncertainty in mapping relationship between computing resources and QoS and varying prices of computing resources across different situations, we propose a diffusion model-based approach to tackle this optimization problem.

\subsubsection{Diffusion Model Generated Optimal Pricing Strategy}
The diffusion model is a type of latent variable model, which first introduces Gaussian noise to perturb training samples, and then learns to perform the inverse denoising process to generate samples similar to the original. This denoising process allows the model to understand the underlying structure of the data, leading to more accurate and realistic generations~\cite{du2023ai}. Hence, we leverage the inverse diffusion process to generate optimal pricing strategy to solve this optimization problem~\cite{du2023beyond}.

Specifically, the forward process of the diffusion model is defined as a Markov chain, wherein $T$ rounds of noises are sequentially added to the training samples. As $T$ approaches infinity, the original samples converge to standard Gaussian noise distribution. For a given distribution ${s_0}$, this forward process can be expressed as follows
\begin{align}\label{eq24}
z\left( {{s_{1:T}}|{s_0}} \right) &= \prod\limits_{t = 1}^T {z\left( {{s_t}|{s_{t - 1}}} \right)} \\ \notag
&= \prod\limits_{t = 1}^T {{{\cal N}}\left( {{s_t};\sqrt {1 - {\beta _t}} {s_{t - 1}},{\beta _t}{\bf{I}}} \right)}, 
\end{align}
where ${\left\{ \beta  \right\}_{t = 1:T}}$ is the hyperparameter corresponding to the variance of Gaussian distribution, ${\bf{I}}$ is the identity matrix. Therefore, for given ${s_0}$, ${s_t}$ can be denoted as
\begin{align}\label{eq25}
z\left( {{s_t}|{s_0}} \right) = {{\cal N}}\left( {{s_t};\sqrt {{{\bar \vartheta }_t}} {s_0},\left( {1 - {{\bar \vartheta }_t}} \right){\bf{I}}} \right),
\end{align}
where ${\bar \vartheta _t} = \prod\nolimits_{i = 1}^t {\left( {1 - {\beta _i}} \right)} $. In contrast to the forward process, the inference stage involves an inverse denoising process to generate samples. Theoretically, if $z\left( {{s_{t - 1}}|{s_t}} \right)$ can be obtained, we can use it to recover the original sample from the standard Gaussian distribution. However, the acquisition of $z\left( {{s_{t - 1}}|{s_t}} \right)$ requires knowledge of all pricing strategies in all conditions, which is difficult to acheive in WiPe-GAI. Therefore, a neural network is used to learn the following transition relation as follows:
\begin{align}\label{eq26}
{p_\omega }\left( {{s_{t - 1}}|{s_t}} \right) = {{\cal N}}\left( {{s_{t - 1}};\mu \left( {{s_t},t} \right),\sigma _\omega ^2\left( {{s_t},t} \right){\bf{I}}} \right),
\end{align}
where $\omega $ is the hyperparameter of the neural network. On this basis, the inverse denoising process can be described as 
\begin{align}\label{eq27}
{p_\omega }\left( {{s_{0:T}}} \right) &= p\left( {{s_T}} \right)\prod\limits_{t = T}^1 {{p_\omega }\left( {{s_{t - 1}}|{s_t}} \right)} \\ \notag
&= p\left( {{s_T}} \right)\prod\limits_{t = T}^1 {{{\cal N}}\left( {{s_{t - 1}};{\mu _\omega }\left( {{s_t},t} \right),\sigma _\omega ^2\left( {{s_t},t} \right){\bf{I}}} \right)}, 
\end{align}
where $p\left( {{s_T}} \right) = {{\cal N}}\left( {{s_T};{\bf{0}},{\bf{I}}} \right)$. As it can be seen, the purpose of training the neural network is to enable it to learn ${\mu _\omega }\left( {{s_t},t} \right)$ and $\sigma _\omega ^2\left( {{s_t},t} \right)$, respectively. From another perspective, given ${s_0}$, the Bayes equation can be utilized to obtain 
\begin{align}\label{eq28}
z\left( {{s_{t - 1}}|{s_t},{s_0}} \right) = {{\cal N}}\left( {{s_{t - 1}};{{\tilde \mu }_t}\left( {{s_t}} \right),{{\tilde \beta }_t}{\bf{I}}} \right),
\end{align}
where ${\tilde \mu _t}\left( {{s_t}} \right) = {{\left( {{s_t} - {{{\beta _t}\bar \varepsilon } \mathord{\left/
 {\vphantom {{{\beta _t}\bar \varepsilon } {\sqrt {1 - {{\bar \vartheta }_t}} }}} \right.
 \kern-\nulldelimiterspace} {\sqrt {1 - {{\bar \vartheta }_t}} }}} \right)} \mathord{\left/
 {\vphantom {{\left( {{s_t} - {{{\beta _t}\bar \varepsilon } \mathord{\left/
 {\vphantom {{{\beta _t}\bar \varepsilon } {\sqrt {1 - {{\bar \vartheta }_t}} }}} \right.
 \kern-\nulldelimiterspace} {\sqrt {1 - {{\bar \vartheta }_t}} }}} \right)} {\sqrt {{\vartheta _t}} }}} \right.
 \kern-\nulldelimiterspace} {\sqrt {{\vartheta _t}} }}$ and ${\tilde \beta _t} = {{\left( {1 - {{\bar \vartheta }_{t{\rm{ - }}1}}} \right){\beta _t}} \mathord{\left/
 {\vphantom {{\left( {1 - {{\bar \vartheta }_{t{\rm{ - }}1}}} \right){\beta _t}} {\left( {1 - {{\bar \vartheta }_t}} \right)}}} \right.
 \kern-\nulldelimiterspace} {\left( {1 - {{\bar \vartheta }_t}} \right)}}$. Considering ${\tilde \mu _t}\left( {{s_t}} \right)$ as the ground truth, therefore, the learned ${\mu _\omega }\left( {{s_t},t} \right)$ is essentially ${\varepsilon _\omega }\left( {{s_t},t} \right)$, due to the relation
\begin{align}\label{eq29}
{\tilde \mu _\omega }\left( {{s_t},t} \right) = \frac{1}{{\sqrt {{\vartheta _t}} }}\left[ {{s_t} - \frac{{{\beta _t}}}{{\sqrt {1 - {{\bar \vartheta }_t}} }}{{\bar \varepsilon }_\omega }\left( {{s_t},t} \right)} \right],
\end{align}
and the prediction result of the model at step $t-1$ is
\begin{align}\label{eq30}
{s_{t - 1}}\left( {{s_t},t;\omega } \right) = \frac{1}{{\sqrt {{\vartheta _t}} }}\left[ {{s_t} - \frac{{{\beta _t}}}{{\sqrt {1 - {{\bar \vartheta }_t}} }}{\varepsilon _\omega }\left( {{s_t},t} \right)} \right] + {\sigma _\omega }\left( {{s_t},t} \right),
\end{align}
where $z \sim{{\cal N}}\left( {{\bf{0}},{\bf{I}}} \right)$. 

Building upon the aforementioned model, and taking into account the influence of parameters such as the cost of computing resources on pricing, we construct a conditional diffusion model and utilize its inverse process to generate the optimal pricing strategy. Specifically, assuming the pricing to be generated is represented by ${\bf{s}} = \left\{ {{v_r},{I_b}} \right\}$, and the state parameters influencing the resource allocation and QoS of the VSP are denoted by ${\bf{c}} = \left\{ {{c_{{\varsigma _s}}},{c_{{\varsigma _{brq}}}},{c_{{\varsigma _{tv}}}},{v_{c}},{v_{r}}, {v_{m}}} \right\}$, then the inverse process of the conditional diffusion model is
\begin{align}\label{eq31}
{p'_\omega }\left( {{\bf{s}}|{\bf{c}}} \right) = {{\cal N}}\left( {{{\bf{s}}^T};{\bf{0}},{\bf{I}}} \right)\prod\limits_{t = T}^1 {{{p'}_\omega }\left( {{{\bf{s}}_{t - 1}}|{{\bf{s}}_{t,}}{\bf{c}}} \right)},
\end{align}
where ${p'_\omega }\left( {{{\bf{s}}_{t - 1}}|{{\bf{s}}_{t,}}{\bf{c}}} \right)$ can be model as a Gaussian distribution expressed as ${{\cal N}}\left( {{{\bf{s}}_{t - 1}};{\mu _\omega }\left( {{{\bf{s}}_t},t,{\bf{c}}} \right),\sigma _\omega ^2\left( {{{\bf{s}}_t},t,{\bf{c}}} \right){\bf{I}}} \right)$, and its corresponding mean and variance are denoted as 
\begin{align}\label{eq32}
\left\{ \begin{array}{l}
{\mu _\omega }\left( {{{\bf{s}}_t},t,{\bf{c}}} \right){\rm{ = }}\frac{1}{{\sqrt {{\vartheta _t}} }}\left[ {{{\bf{s}}_t} - \frac{{{\beta _t}}}{{\sqrt {1 - \sqrt {{{\bar \vartheta }_t}} } }}{\varepsilon _\omega }\left( {{{\bf{s}}_t},t,{\bf{c}}} \right)} \right],\\
\sigma _\omega ^2\left( {{{\bf{s}}_t},t,{\bf{c}}} \right) = {\beta _t}{\bf{I}},
\end{array}\right.
\end{align}
respectively. Meanwhile, according to~\eqref{eq31}, under the condition of ${\bf{c}}$, the prediction outcome of the conditional diffusion model inverse process at step $t-1$ can be expressed as
\begin{align}\label{eq33}
{{\bf{s}}_{t - 1}}\left( {{{\bf{s}}_t},t,{\bf{c}};\omega } \right) &= \frac{1}{{\sqrt {{\vartheta _t}} }}\left[ {{{\bf{s}}_t} - \frac{{{\beta _t}}}{{\sqrt {1 - {{\bar \vartheta }_t}} }}{\varepsilon _\omega }\left( {{{\bf{s}}_t},t,{\bf{c}}} \right)} \right] \\ \notag
&+ {\sigma _\omega }\left( {{{\bf{s}}_t},t,{\bf{c}}} \right)\varepsilon.
\end{align}
In WiPe-GAI, our training objective is to determine an ${\varepsilon _\omega }$ capable of generating an optimal ${\bf{s}}_0$ given the condition ${\bf c}$. Here, an optimal ${\bf{s}}_0$ is defined as one that maximizes ${U_{us}}$ subject to the constraints defined in \eqref{eq23}. Drawing inspiration from the deep reinforcement learning paradigm, we redefine certain elements in our context. Here, ${\bf{c}}$ is treated as the environment, while ${\bf{s}}_0$ is considered the action. The expected cumulative reward is represented as the Q-value, denoted as $Q_v({\bf{s}}_0, {\bf{c}})$.
To manage the training process, Q-learning is adopted. Hence, the optimal ${\varepsilon _\omega }$ becomes synonymous with a denoising network that maximizes the expected cumulative Q-values, which can be expressed as
\begin{equation}\label{eq34}
\mathop {\arg \min }\limits_{{\varepsilon _\omega }} \mathcal{L}(\omega  ) =  - {\mathbb{E}_{{{\mathbf{s}}_0}\sim{\varepsilon _\omega }}}\left[ {{Q_v}\left( {{{\mathbf{s}}_0},{\mathbf{c}}} \right)} \right].
\end{equation}
Upon completion of the training, the resulting model is utilized to generate the optimal strategy, by solving the optimization problem in~\eqref{eq23}. The overall training and inference process is summarized in Algorithm 1.
 
\begin{algorithm}[t]
{\small \caption{Diffusion Model Generated Optimal Pricing Strategy}}
\hspace*{0.02in} {\bf{\textit{Training Phase:}}}
\begin{algorithmic}[1]
\State Input hyper-parameters: denoising step $T$, initialize neural network parameters $\omega $ and $v$
\vspace{0.1cm}
\State \#\#{\textit{\quad Learning Process}}
\State Initialize a random process for pricing strategy exploration
\While{not converge}
\State Observe the current environment \\ \qquad\qquad${\bf{c}} = \left\{ {{c_{{\varsigma _s}}},{c_{{\varsigma _{brq}}}},{c_{{\varsigma _{tv}}}},{v_{c}},{v_{r}}, {v_{m}}} \right\}$
\State Set ${\bm{s}}_N$ as Gaussian noise. Generate pricing strategy ${\bm{s}}_0$ by denoising ${\bm{s}}_N$ according to~\eqref{eq33}
\State Apply the generated pricing strategy ${\bm{s}}_0$ to the environment and observe the utility value as~\eqref{eq22}.
\State Record the real utility value
\State Update $Q_v$ by minimizing the mean squared error between the real and predicted utility values
\State Update $\varepsilon_\omega $ according to~\eqref{eq34}
\EndWhile
\State \Return The trained solution generation network ${\bm{\varepsilon}}_\theta$
\end{algorithmic}
\hspace*{0.02in} {\bf{\textit{Inference Phase:}}}
\begin{algorithmic}[1]
\State Observe the environment vector ${\bm c}$
\State Generate the optimal pricing strategy ${\bm{s}}_0$ by denoising Gaussian noise using ${\bm{\varepsilon}}_\theta$
\State \Return The optimal pricing strategy ${\bm s}_0$
\label{Algorithm1}
\end{algorithmic}
\end{algorithm}

\section{Experiment and Evaluation}
In this section, we conduct a comprehensive evaluation and analysis of the proposed WiPe-GAI framework through experiments from two perspectives. First, we evaluate the performance of the user skeleton extraction and virtual character generation, based on collected CSI data, to validate the feasibility of WiPe-GAI. Utilizing the evaluation results, then, we obtain the mapping functions ${\varsigma _s}\left(  \cdot  \right)$, ${\varsigma _{brq}}\left(  \cdot  \right)$, and ${\varsigma _{tv}}\left(  \cdot  \right)$ through fitting and perform experiments to evaluate the efficiency of the proposed incentive mechanisms. 
\subsection{Experimental Configuration}
In the experiments, multiple APs equipped with the Broadcom 4366C0 chips and the Nexmon toolkit~\cite{gringoli2019free} are used to collect CSI data based on the IEEE 802.11ac protocol. The AP operates at 5.805 GHz with the signal bandwidth of 80 MHz (including 256 subcarriers) and the transmission rate of 100 packets per second. During the perception process, the transmitter utilizes a single antenna for signal transmission and the receiver employs four antennas to collect CSI, while one of them is used for phase error cancellation and the others for user skeleton generation. The proposed algorithms are executed on an experimental platform constructed on a standard Ubuntu 20.04 system, equipped with an AMD Ryzen Threadripper PRO 3975WX 32-core processor and an NVIDIA RTX A5000 graphics processing unit (GPU).
\subsection{Wireless Perception to Virtual Character Generation}
\subsubsection{Effectiveness of WiPe-GAI} 
\begin{figure*}[t]
	\centering
	\includegraphics[width=0.9\textwidth]{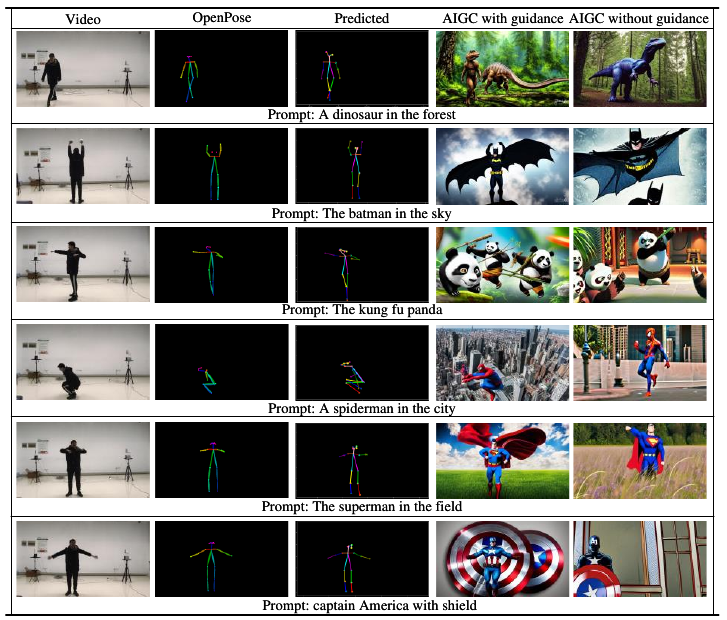}%
	\caption{The predicted user skeleton and the generated virtual character. In the figures, the first column presents the user's posture captured by camera in the real-world scenario. The second column depicts the posture predicted by OpenPose based on the captured video sequence. The third and fourth columns, respectively, illustrate the user's skeleton predicted by WiPe-GAI and the generated corresponding virtual character. The fifth column displays the virtual characters generated without perceptual guidance.}
	\label{VOPA}
\end{figure*}

To verify the effectiveness of WiPe-GAI, we first conduct experiments on user skeleton prediction and the virtual character generation, the results are presented in Fig.~\ref{VOPA}. Taking the skeleton predicted by OpenPose~\cite{cao2021openpose} as the ground truth, from the figures, we can observe that WiPe-GAI can effectively predict the skeleton of the user via CSI by using the proposed SMSP algorithm and the ${\bf{S}}\left(  \cdot  \right)$. There are some differences between our predicted skeleton and the user's actual posture. For instance, some differences exist between the positions of the predicted knee and the real knee of the user, as can be seen in the second row of results. However, these differences are small and the overall skeleton extracted from CSI is fairly close to the user's real posture. This validates the effectiveness of the proposed SMSP based skeleton extraction. 
Building on this, the predicted skeleton and the user's prompts are used as external conditions and prompts, respectively, to generate the virtual character for the user. As can be seen from the fourth column in Fig.~\ref{VOPA}, WiPe-GAI is able to effectively generate the virtual character based on the predicted skeleton and user's prompt. Compared to the results in the fifth column without perception guidance, WiPe-GAI produces a virtual character whose posture aligns more accurately with the user's actual posture, demonstrating the effectiveness of the proposed WiPe-GAI framework. Furthermore, WiPe-GAI can craft a fitting background for the virtual character based on user's prompts, thereby enhancing the naturalness of the overall generated image.

\subsubsection{Impact of AP Quantity on Skeleton Prediction} 
\begin{figure*}[t]
	\centering
\includegraphics[width=0.9\textwidth]{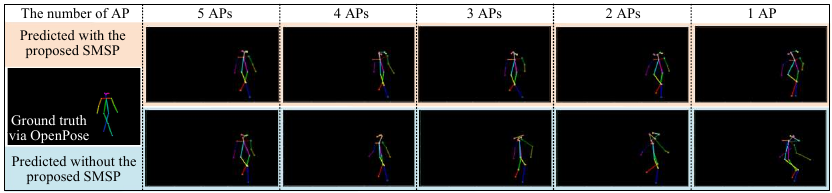}%
	\caption{The impact of AP quantity on skeleton prediction.}
	\label{APImPCT}
\end{figure*}

After verifying the effectiveness of WiPe-GAI, we next analyze the effect of the number of APs on perception accuracy, and compare our approach with the existing method in~\cite{wang2019can}. The results are presented in Fig.~\ref{APImPCT}. As the results show, the skeleton prediction performance deteriorates as the number of APs decreases. This can be explained by the fact that a decrease in the AP quantity causes a reduction in the information about user posture contained in the CSI feature matrix, which subsequently leads to a drop in prediction accuracy. However, given the fixed total computational resource, using fewer APs would free up more resources for the GAI, which can enhance the AIGC quality.

Furthermore, a comparison between the results in the first and second rows reveals that the skeleton prediction accuracy of the proposed SMSP algorithm outperforms the methods that directly use the original CSI data for skeleton prediction~\cite{wang2019can}, especially when fewer APs are involved. For instance, with perception involves only one AP, the skeleton predicted by our algorithm can roughly indicate that the user is in a standing position, while the prediction of~\cite{wang2019can} implies that the user is in a squatting position, which does not match the ground truth.

\begin{figure*}[t]
\centering
\includegraphics[width=0.9\textwidth]{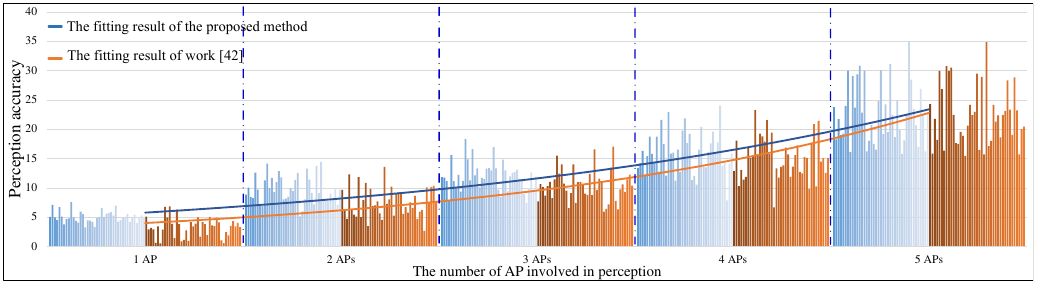} 
\caption{The relation between the number of APs involved in perception and the perception performance.} 
\label{P_RLT} 
\end{figure*}
Using the reciprocal of the normalized Euclidean distance between the predicted skeleton and the skeleton obtained by OpenPose as metric, we conduct multiple predictions under different numbers of APs and analyze the prediction accuracy. The results are shown in Fig.~\ref{P_RLT}. For the proposed WiPe-GAI framework and the method in~\cite{wang2019can}, as shown by the blue and red bars in the figure, respectively, the more APs, the more user information the feature CSI matrix contains, thereby resulting in a higher prediction accuracy for the skeleton. Specifically, with one AP involved, the prediction accuracy of WiPe-GAI and the method in~\cite{wang2019can} approximates 5.7 and 4.2, respectively. However, as the number of APs increases to 5, the prediction accuracy of these two methods improves to around 23.5 and 22.9, respectively. At the same time, as the number of APs increases, it can be found that the performance of the method in~\cite{wang2019can} gradually approaches to that of WiPe-GAI. This is because an increase in the number of APs results in an additional amount of information related to the user's posture, enabling a more accurate prediction even without specific signal processing.

By fitting the prediction results of both systems, the mapping relationships between the number of APs and the perception accuracy can be obtained, as shown by the red and blue lines in Fig.~\ref{P_RLT}. From the fitting results, it can be seen that the overall prediction performance of WiPe-GAI is better, especially with fewer APs, demonstrating the efficacy of the proposed SMSP algorithm. Essentially, the obtained mapping relationship signifies the relationship between computational resources and perception accuracy, since the more APs involved in perception, the higher the resource consumption for prediction. Therefore, we use the fitted relationship as ${\varsigma _s}\left( \cdot \right)$ for the following analysis.

\subsubsection{Impact of Inference Steps on Virtual Character Generation} 
\begin{figure*}[t]
	\centering
	\includegraphics[width=0.9\textwidth]{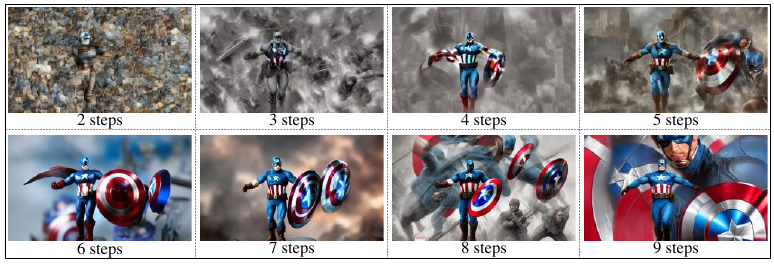}%
	\caption{Impact of inference steps on virtual character generation.}
	\label{STEPIPCT}
\end{figure*}

In addition to the perception, we also analyze the impact of the number of inference steps on the generation of virtual characters, the results are illustrated in Fig.~\ref{STEPIPCT}. From the figures, it is clear that the quality of the generated virtual character improves as the number of inference steps increases. Specifically, the virtual character generated with only 2 to 3 inference steps are predominantly in black and white, with incomplete character limbs, as the first two figures in the results show. However, with more inference steps, these issues are effectively alleviated, exhibiting a character with more thematic color, complete limbs, and less noise in the background. This is understandable, as more steps implies that the GAI model can perform more in-depth denoising, thereby producing higher quality results.

\begin{figure}[t]
\centering
\includegraphics[width=0.45\textwidth]{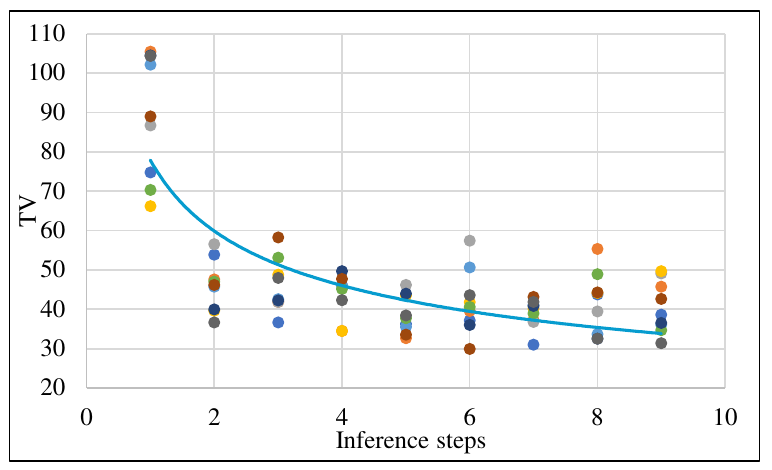} 
\caption{The TV value versus the number of inference steps.} 
\label{STEPTV} 
\end{figure}

\begin{figure}[t]
\centering
\includegraphics[width=0.45\textwidth]{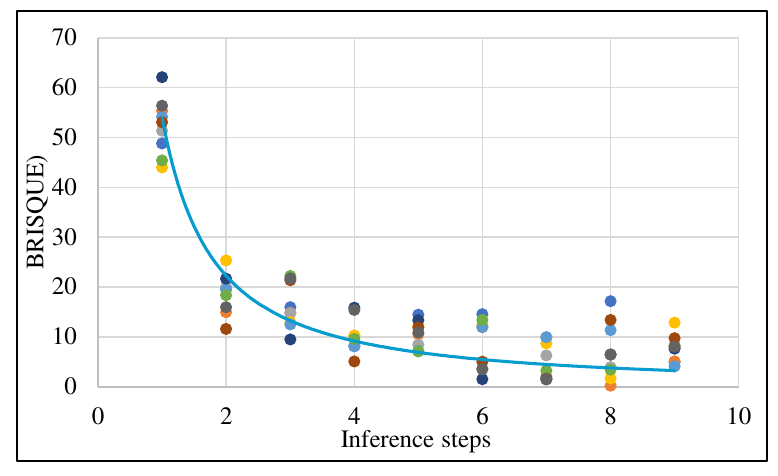} 
\caption{The BRISQUE value versus the number of inference steps.} 
\label{STEPBRI} 
\end{figure}

On this basis, we further calculate the BRISQUE and TV values based on the images generated from multiple experiments. The results, represented as data points, are shown in Fig.~\ref{STEPTV} and Fig.~\ref{STEPBRI}. According to the results, we observe a decrease in the TV value (from around 78 to 32) and a drop in the BRISQUE value (from approximately 55 to 3), as the number of inference steps increases from 1 to 10. These decreasing trends suggest an improvement in the naturalness and smoothness of the generated image, which contains the virtual character and the corresponding background, while also showing that GAI consumes more resources. By fitting these data points, we obtain the relationship between the computation resources allocated to GAI and the quality of the generated digital content, as the blue curves show. Hence, we use them as ${\varsigma _{tv}}\left( \cdot \right)$ and ${\varsigma _{brq}}\left( \cdot \right)$ for subsequent analysis.

\subsection{Incentive Mechanism Analysis} 
\subsubsection{Pricing Strategy Generation} Using the obtained ${\varsigma _s}\left( \cdot \right)$, ${\varsigma _{tv}}\left( \cdot \right)$, and ${\varsigma _{brq}}\left( \cdot \right)$, we analyze the optimal pricing strategy generated by diffusion model and compare it with two deep reinforcement learning (DRL) algorithms, i.e., Soft Actor-Critic (SAC)~\cite{haarnoja2018soft} and Proximal Policy Optimization (PPO)~\cite{schulman2017proximal}. The PPO realizes optimization by using a clipped surrogate objective to update the policy iteratively, which can provide smooth policy changes. The SAC is an off-policy algorithm, which maximizes the expected cumulative reward and the entropy of the policy by learning a stochastic policy. During the experiments, we assume that the VSP has a maximum of 100 units of computational resources, with the processing of CSI data of a single AP consuming 2 units, the prediction of the skeleton requiring 1 unit, and each inference step using 2 units.

\begin{figure}[t]
\centering
\includegraphics[width=0.5\textwidth]{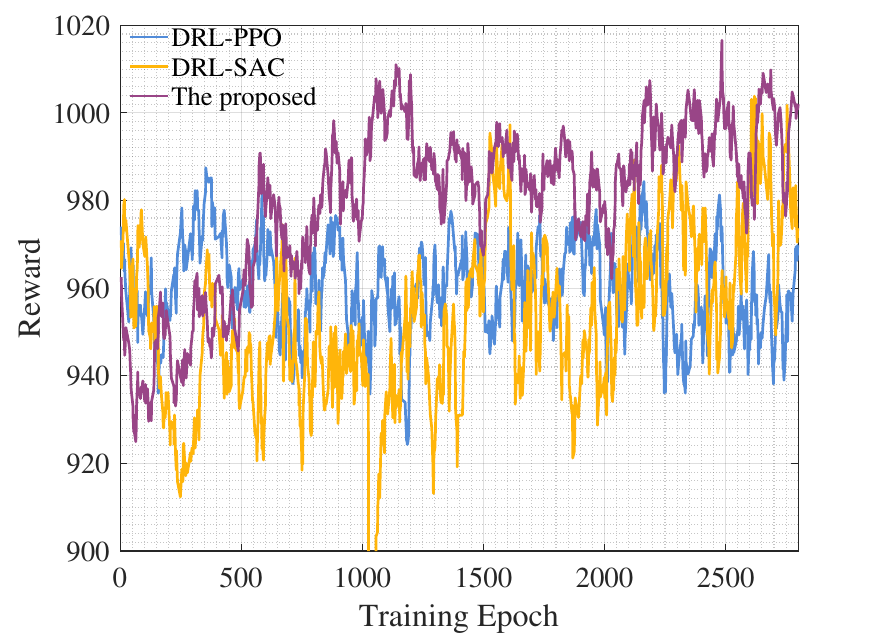} 
\caption{The training curves, with the diffusion step of 10, batch size of 512, soft target update parameter of 0.005, discount factor of 0.95, exploration noise of 0.01, and learning rate of ${10^{-5}}$.} 
\label{TRAIN} 
\end{figure}

The results in Fig.~\ref{TRAIN} show the achievable reward against the training epoch of the proposed algorithm in comparison with SAC and PPO. From the experimental results, it can be observed that, under the preset number of epochs, the proposed algorithm has already converged, while SAC and PPO do not show a clear trend of convergence, indicating that the proposed algorithm converges faster. Moreover, the reward of the proposed algorithm is about 1000, whereas DRL-SAC and DRL-PPO can achieve around 970 and 960, respectively, which is lower than that of the proposed algorithm. We believe this is due to two main reasons. First, the proposed algorithm has a better sampling quality, as the diffusion model can reduce the influence of uncertainty and noise through multiple rounds of fine-tuning. Second, unlike traditional neural networks that only consider the input at the current time step, the diffusion model can generate samples for more time steps by fine-tuning, providing a stronger processing capability for tasks with long-term dependencies.

\begin{figure}[t]
\centering
\includegraphics[width=0.45\textwidth]{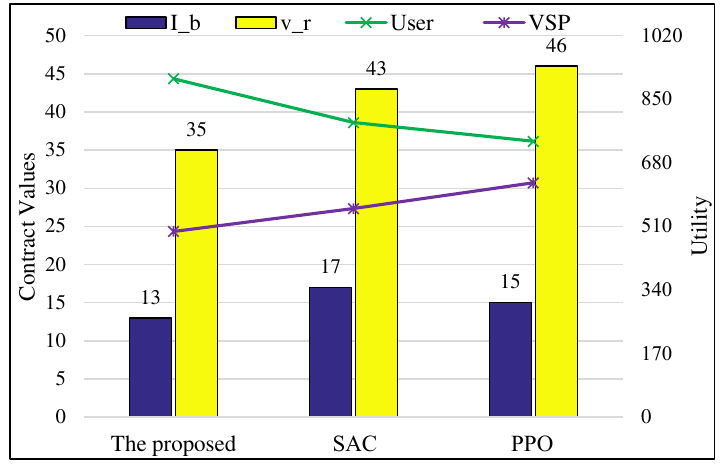} 
\caption{The generated optimal pricing strategy and the corresponding utility of user and the VSP.} 
\label{UTLT} 
\end{figure}

Using the trained models, we further compare the optimal pricing strategy design capabilities of different models under a given environment state. The results of this comparison are presented in Fig.~\ref{UTLT}. As can be seen from the figure, the strategy generated by the proposed method (with ${I_b}=13$ and ${v_r}=35$) yields a user utility of 910, exceeding the utility of 787 and 737, which are achieved by DRL-SAC (with ${I_b}=17$ and ${v_r}=43$) and DRL-PPO (with ${I_b}=15$ and ${v_r}=46$), respectively. A noteworthy detail is that the VSP's utility provided by the optimal pricing strategy generated by the diffusion model stands at 496, which is lower than 557 and 626 achieved by SAC and PPO, respectively. We believe that this trade-off is reasonable, as the pricing strategy aims to maximize the utility of the user while still incentivizing the VSP's participation.

\subsubsection{Impact of Perception on Incentive Mechanism}
\begin{figure}[t]
\centering
\includegraphics[width=0.48\textwidth]{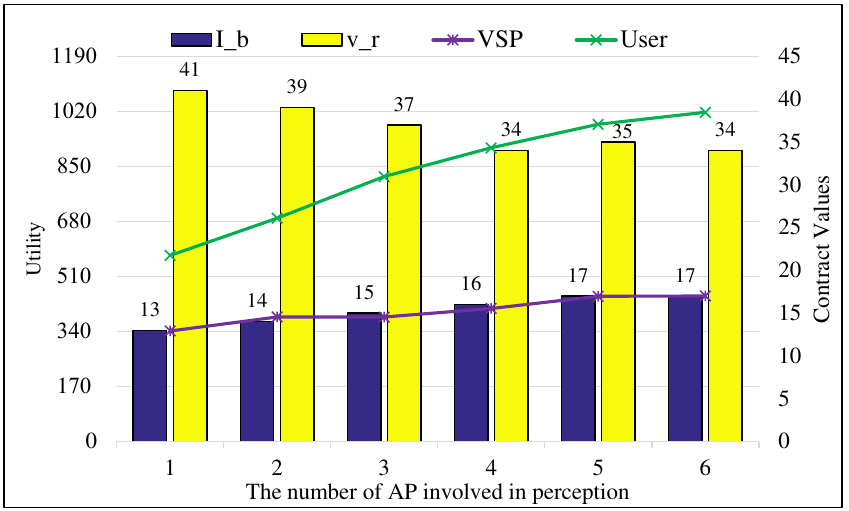} 
\caption{The impact of the number of APs involved in perception on the utility of the user and VSP.} 
\label{APIPCT} 
\end{figure}

In some practical scenarios, the number of APs available for perception in the physical environment may be relatively limited. Hence, we analyze the influence of the number of APs on the incentive mechanism. The results are presented in Fig.~\ref{APIPCT}. As can be seen, when the total number of APs is relatively small, an increase in the number of APs yields an enhancement in the utility of both the user and VSP, while ${v_r}$ and the total amount that the user needs to pay are both decreasing. Specifically, when a single AP is involved in perception, the generated optimal pricing strategy is $\left({I_b}=13, {v_r}=41\right)$, and the utility of the user and the VSP are 575 and 341, respectively. However, when the perception incorporates 6 APs, ${I_b}$ increases to 17, ${v_r}$ falls to 34, and the utility of user and the VSP increase to 1016 and 450, respectively. 

This is because, when there are few APs involved in perception, the QoS of perception (i.e., ${Q_s}$) is low, driving the VSP to allocate more resources to the GAI. The aim of WiPe-GAI adopting this strategy is to enhance ${Q_t}$ by increasing the number of inference steps, so as to maximize the VSP's utility and guarantee its participation in service provisioning. However, once the number of inference steps reaches a certain level, the rate of increase in ${Q_t}$ slows down, which forces the user to further increase ${v_r}$ to ensure the VSP's participation in service provision. Fortunately, as the number of APs gradually rises, the QoS improvement brought about by perception exceeds that of AIGC when consuming unit energy. Consequently, the VSP reassigns some of the resources initially allocated to GAI to perception, therefore maximizing its utility and ensuring its participation in service provision. From another perspective, this reallocation strategy not only reduces ${v_r}$ but also enhances the user utility, verifying the rationality of the generated optimal pricing strategy and further illustrates the effectiveness of the proposed framework.

\subsection{Discussion}
In the experiments presented above, we conduct a comprehensive evaluation of the proposed WiPe-GAI framework from perspectives of skeleton prediction, virtual character generation, and incentive mechanism. From these results, we can observe the following critical points:
\begin{itemize}
\item The proposed SMSP algorithm utilizes the information about user posture contained within CSI more effectively, thus enhancing the performance of user skeleton prediction and, overall, outperforming the method without SMSP.
\item Using the predicted skeleton and user's requests, i.e., prompts, WiPe-GAI can effectively generate the virtual character and the corresponding background for the user, verifying the effectiveness of the proposed framework.

\item The proposed diffusion model based method can efficiently generate the optimal pricing strategy, better than the conventional DRL based methods in terms of maximizing the user's utility and speed of convergence, while also encouraging the VSP to actively participate in service provision.
\end{itemize}
Besides these achievements, the proposed WiPe-GAI has certain limitations, which are summarized as follows:
\begin{itemize}
\item The proposed SMSP improves the performance of CSI-based skeleton prediction, but it may show unsatisfactory results when fewer APs are available. One possible solution to address this issue is to optimize the deployment of APs, so that each AP can collect more non-overlapping information at different spatial locations for prediction .
\item This paper only uses image as examples of the generated digital content. Yet, practical applications may require video streams to be produced for users. Given that video generation demands more resources, retraining the model might be necessary. However, the framework and optimization strategies proposed in this paper should remain effective.
\item While the proposed diffusion based model demonstrates faster convergence in optimal pricing strategy generation, each execution involves a multi-step denoising process, which may not be outstanding in terms of efficiency. Considering the complexity of real-world applications, furthe refining the efficiency of the proposed model is needed.
\end{itemize}



\section{Conclusion}
This paper introduces WiPe-GAI, a framework that combines wireless perception with GAI to provide the AIGC service to users. For WiPe-GAI, we introduced a novel SMSP algorithm, which uses CSI to predict the user's skeleton, thereby guiding the GAI to generate virtual characters for the user. Furthermore, to encourage the VSP to participate in service provision, WiPe-GAI builds an incentive mechanism based on pricing and incorporates an new diffusion-based method to generate optimal pricing strategy, which maximizes user's utility while ensuring the VSP's participation. Through comprehensive experiments, it was demonstrated that WiPe-GAI can accurately predict the user's skeleton and generate the corresponding virtual character for the user. Furthermore, the proposed diffusion-based approach can effectively generate the optimal pricing strategy, which not only yields greater user utility, but also ensures that the VSP’s participation, outperforming the existing DRL based methods. For future work, we plan to refine the proposed framework by incorporating additional factors such as communication loss and multi-user concurrency. Meanwhile, we will continue  to explore the application of optimization methods based on the diffusion model in various domains.
\bibliographystyle{IEEEtran}
\bibliography{Ref.bib} 

\begin{thebibliography}{10}
\providecommand{\url}[1]{#1}
\csname url@samestyle\endcsname
\providecommand{\newblock}{\relax}
\providecommand{\bibinfo}[2]{#2}
\providecommand{\BIBentrySTDinterwordspacing}{\spaceskip=0pt\relax}
\providecommand{\BIBentryALTinterwordstretchfactor}{4}
\providecommand{\BIBentryALTinterwordspacing}{\spaceskip=\fontdimen2\font plus
\BIBentryALTinterwordstretchfactor\fontdimen3\font minus
  \fontdimen4\font\relax}
\providecommand{\BIBforeignlanguage}[2]{{%
\expandafter\ifx\csname l@#1\endcsname\relax
\typeout{** WARNING: IEEEtran.bst: No hyphenation pattern has been}%
\typeout{** loaded for the language `#1'. Using the pattern for}%
\typeout{** the default language instead.}%
\else
\language=\csname l@#1\endcsname
\fi
#2}}
\providecommand{\BIBdecl}{\relax}
\BIBdecl

\bibitem{xu2023unleashing}
M.~Xu, H.~Du, D.~Niyato, J.~Kang, Z.~Xiong, S.~Mao, Z.~Han, A.~Jamalipour,
  D.~I. Kim, X.~Shen, V.~Leung, and P.~H.~Vincent, ``Unleashing the power of
  edge-cloud generative ai in mobile networks: A survey of aigc services,''
  \emph{arXiv preprint arXiv:2303.16129}, 2023.

\bibitem{koksal2023controllable}
A.~K{\"o}ksal, K.~E. Ak, Y.~Sun, D.~Rajan, and J.~H. Lim, ``Controllable video
  generation with text-based instructions,'' \emph{IEEE Transactions on
  Multimedia}, 2023.

\bibitem{wu2023ai}
J.~Wu, W.~Gan, Z.~Chen, S.~Wan, and H.~Lin, ``Ai-generated content (aigc): A
  survey,'' \emph{arXiv preprint arXiv:2304.06632}, 2023.

\bibitem{cao2023comprehensive}
Y.~Cao, S.~Li, Y.~Liu, Z.~Yan, Y.~Dai, P.~S. Yu, and L.~Sun, ``A comprehensive
  survey of ai-generated content (aigc): A history of generative ai from gan to
  chatgpt,'' \emph{arXiv preprint arXiv:2303.04226}, 2023.

\bibitem{croitoru2023diffusion}
F.-A. Croitoru, V.~Hondru, R.~T. Ionescu, and M.~Shah, ``Diffusion models in
  vision: A survey,'' \emph{IEEE Transactions on Pattern Analysis and Machine
  Intelligence}, 2023.

\bibitem{zhang2023adding}
L.~Zhang and M.~Agrawala, ``Adding conditional control to text-to-image
  diffusion models,'' \emph{arXiv preprint arXiv:2302.05543}, 2023.

\bibitem{wang2023guiding}
J.~Wang, H.~Du, D.~Niyato, Z.~Xiong, J.~Kang, S.~Mao, and X.~Shen, ``Guiding
  {AI}-generated digital content with wireless perception,'' \emph{arXiv
  preprint arXiv:2303.14624}, 2023.

\bibitem{bond2021deep}
S.~Bond-Taylor, A.~Leach, Y.~Long, and C.~G. Willcocks, ``Deep generative
  modelling: A comparative review of vaes, gans, normalizing flows,
  energy-based and autoregressive models,'' \emph{IEEE transactions on pattern
  analysis and machine intelligence}, 2021.

\bibitem{du2023generative}
H.~Du, Z.~Li, D.~Niyato, J.~Kang, Z.~Xiong, H.~Huang, and S.~Mao, ``Generative
  {AI}-aided optimization for {AI}-generated content (aigc) services in edge
  networks,'' \emph{arXiv preprint arXiv:2303.13052}, 2023.

\bibitem{wang2016csi}
X.~Wang, L.~Gao, S.~Mao, and S.~Pandey, ``{CSI}-based fingerprinting for indoor
  localization: A deep learning approach,'' \emph{IEEE transactions on
  vehicular technology}, vol.~66, no.~1, pp. 763--776, 2016.

\bibitem{tan2020enabling}
S.~Tan, J.~Yang, and Y.~Chen, ``Enabling fine-grained finger gesture
  recognition on commodity {WiFi} devices,'' \emph{IEEE Transactions on Mobile
  Computing}, vol.~21, no.~8, pp. 2789--2802, 2020.

\bibitem{karanam20173d}
C.~R. Karanam and Y.~Mostofi, ``{3D} through-wall imaging with unmanned aerial
  vehicles using wifi,'' in \emph{Proceedings of the 16th ACM/IEEE
  International Conference on Information Processing in Sensor Networks}, 2017,
  pp. 131--142.

\bibitem{zhao2018rf}
M.~Zhao, Y.~Tian, H.~Zhao, M.~A. Alsheikh, T.~Li, R.~Hristov, Z.~Kabelac,
  D.~Katabi, and A.~Torralba, ``{RF}-based {3D} skeletons,'' in
  \emph{Proceedings of the 2018 Conference of the ACM Special Interest Group on
  Data Communication}, 2018, pp. 267--281.

\bibitem{zhao2018through}
M.~Zhao, T.~Li, M.~Abu~Alsheikh, Y.~Tian, H.~Zhao, A.~Torralba, and D.~Katabi,
  ``Through-wall human pose estimation using radio signals,'' in
  \emph{Proceedings of the IEEE Conference on Computer Vision and Pattern
  Recognition}, 2018, pp. 7356--7365.

\bibitem{yang2020rfid}
C.~Yang, X.~Wang, and S.~Mao, ``{RFID}-pose: Vision-aided three-dimensional
  human pose estimation with radio-frequency identification,'' \emph{IEEE
  transactions on reliability}, vol.~70, no.~3, pp. 1218--1231, 2020.

\bibitem{guo2019signal}
L.~Guo, Z.~Lu, X.~Wen, S.~Zhou, and Z.~Han, ``From signal to image: Capturing
  fine-grained human poses with commodity {Wi-Fi},'' \emph{IEEE Communications
  Letters}, vol.~24, no.~4, pp. 802--806, 2019.

\bibitem{zhou2023metafi++}
Y.~Zhou, H.~Huang, S.~Yuan, H.~Zou, L.~Xie, and J.~Yang, ``{MetaFi++}:
  {WiFi}-enabled transformer-based human pose estimation for metaverse avatar
  simulation,'' \emph{IEEE Internet of Things Journal}, 2023.

\bibitem{wang2019person}
F.~Wang, S.~Zhou, S.~Panev, J.~Han, and D.~Huang, ``Person-in-{WiFi}:
  Fine-grained person perception using wifi,'' in \emph{Proceedings of the
  IEEE/CVF International Conference on Computer Vision}, 2019, pp. 5452--5461.

\bibitem{du2023beyond}
H.~Du, R.~Zhang, Y.~Liu, J.~Wang, Y.~Lin, Z.~Li, D.~Niyato, J.~Kang, Z.~Xiong,
  S.~Cui \emph{et~al.}, ``Beyond deep reinforcement learning: {A} tutorial on
  generative diffusion models in network optimization,'' \emph{arXiv preprint
  arXiv:2308.05384}, 2023.

\bibitem{ho2022cascaded}
J.~Ho, C.~Saharia, W.~Chan, D.~J. Fleet, M.~Norouzi, and T.~Salimans,
  ``Cascaded diffusion models for high fidelity image generation.'' \emph{J.
  Mach. Learn. Res.}, vol.~23, no.~47, pp. 1--33, 2022.

\bibitem{ma2023unified}
Y.~Ma, H.~Yang, W.~Wang, J.~Fu, and J.~Liu, ``Unified multi-modal latent
  diffusion for joint subject and text conditional image generation,''
  \emph{arXiv preprint arXiv:2303.09319}, 2023.

\bibitem{rombach2022high}
R.~Rombach, A.~Blattmann, D.~Lorenz, P.~Esser, and B.~Ommer, ``High-resolution
  image synthesis with latent diffusion models,'' in \emph{Proceedings of the
  IEEE/CVF Conference on Computer Vision and Pattern Recognition}, 2022, pp.
  10\,684--10\,695.

\bibitem{cheng2023layoutdiffuse}
J.~Cheng, X.~Liang, X.~Shi, T.~He, T.~Xiao, and M.~Li, ``Layoutdiffuse:
  Adapting foundational diffusion models for layout-to-image generation,''
  \emph{arXiv preprint arXiv:2302.08908}, 2023.

\bibitem{wang2022diffusion}
Z.~Wang, J.~J. Hunt, and M.~Zhou, ``Diffusion policies as an expressive policy
  class for offline reinforcement learning,'' \emph{arXiv preprint
  arXiv:2208.06193}, 2022.

\bibitem{du2023ai}
H.~Du, J.~Wang, D.~Niyato, J.~Kang, Z.~Xiong, and D.~I. Kim, ``{AI}-generated
  incentive mechanism and full-duplex semantic communications for information
  sharing,'' \emph{arXiv preprint arXiv:2303.01896}, 2023.

\bibitem{luong2018applications}
N.~C. Luong, P.~Wang, D.~Niyato, Y.-C. Liang, Z.~Han, and F.~Hou,
  ``Applications of economic and pricing models for resource management in {5G}
  wireless networks: A survey,'' \emph{IEEE Communications Surveys \&
  Tutorials}, vol.~21, no.~4, pp. 3298--3339, 2018.

\bibitem{zhao2020intelligent}
Z.~Zhao, W.~Zhou, D.~Deng, J.~Xia, and L.~Fan, ``Intelligent mobile edge
  computing with pricing in internet of things,'' \emph{IEEE Access}, vol.~8,
  pp. 37\,727--37\,735, 2020.

\bibitem{qian2020multi}
B.~Qian, H.~Zhou, T.~Ma, K.~Yu, Q.~Yu, and X.~Shen, ``Multi-operator spectrum
  sharing for massive iot coexisting in {5G/B5G} wireless networks,''
  \emph{IEEE Journal on Selected Areas in Communications}, vol.~39, no.~3, pp.
  881--895, 2020.

\bibitem{yang2022joint}
Y.~Yang, Z.~Liu, Z.~Liu, K.~Y. Chan, Y.~Xie, and X.~Guan, ``Joint optimization
  of edge computing resource pricing and wireless caching for blockchain-driven
  networks,'' \emph{IEEE Transactions on Vehicular Technology}, vol.~71, no.~6,
  pp. 6661--6670, 2022.

\bibitem{qian2020leveraging}
B.~Qian, H.~Zhou, T.~Ma, Y.~Xu, K.~Yu, X.~Shen, and F.~Hou, ``Leveraging
  dynamic stackelberg pricing game for multi-mode spectrum sharing in
  {5G-VANET},'' \emph{IEEE Transactions on Vehicular Technology}, vol.~69,
  no.~6, pp. 6374--6387, 2020.

\bibitem{yang2013rssi}
Z.~Yang, Z.~Zhou, and Y.~Liu, ``From rssi to csi: Indoor localization via
  channel response,'' \emph{ACM Computing Surveys (CSUR)}, vol.~46, no.~2, pp.
  1--32, 2013.

\bibitem{kotaru2015spotfi}
M.~Kotaru, K.~Joshi, D.~Bharadia, and S.~Katti, ``Spotfi: Decimeter level
  localization using wifi,'' in \emph{Proceedings of the 2015 ACM Conference on
  Special Interest Group on Data Communication}, 2015, pp. 269--282.

\bibitem{wax1985detection}
M.~Wax and T.~Kailath, ``Detection of signals by information theoretic
  criteria,'' \emph{IEEE Transactions on acoustics, speech, and signal
  processing}, vol.~33, no.~2, pp. 387--392, 1985.

\bibitem{vasisht2016decimeter}
D.~Vasisht, S.~Kumar, and D.~Katabi, ``{Decimeter-Level} localization with a
  single {WiFi} access point,'' in \emph{13th USENIX Symposium on Networked
  Systems Design and Implementation (NSDI 16)}, 2016, pp. 165--178.

\bibitem{zhang2021fresnel}
D.~Zhang, F.~Zhang, D.~Wu, J.~Xiong, and K.~Niu, ``Fresnel zone based theories
  for contactless sensing,'' \emph{Contactless Human Activity Analysis}, pp.
  145--164, 2021.

\bibitem{zeng2021exploring}
Y.~Zeng, J.~Liu, J.~Xiong, Z.~Liu, D.~Wu, and D.~Zhang, ``Exploring multiple
  antennas for long-range {WiFi} sensing,'' \emph{Proceedings of the ACM on
  Interactive, Mobile, Wearable and Ubiquitous Technologies}, vol.~5, no.~4,
  pp. 1--30, 2021.

\bibitem{cao2021openpose}
Z.~Cao, G.~Hidalgo, T.~Simon, S.-E. Wei, and Y.~Sheikh, ``{OpenPose}: realtime
  multi-person 2d pose estimation using part affinity fields,'' \emph{IEEE
  transactions on pattern analysis and machine intelligence}, vol.~43, no.~1,
  pp. 172--186, 2021.

\bibitem{he2016deep}
K.~He, X.~Zhang, S.~Ren, and J.~Sun, ``Deep residual learning for image
  recognition,'' in \emph{Proceedings of the IEEE conference on computer vision
  and pattern recognition}, 2016, pp. 770--778.

\bibitem{ioffe2015batch}
S.~Ioffe and C.~Szegedy, ``Batch normalization: Accelerating deep network
  training by reducing internal covariate shift,'' in \emph{International
  conference on machine learning}.\hskip 1em plus 0.5em minus 0.4em\relax pmlr,
  2015, pp. 448--456.

\bibitem{krizhevsky2017imagenet}
A.~Krizhevsky, I.~Sutskever, and G.~E. Hinton, ``Imagenet classification with
  deep convolutional neural networks,'' \emph{Communications of the ACM},
  vol.~60, no.~6, pp. 84--90, 2017.

\bibitem{li2023blip}
D.~Li, J.~Li, and S.~C. Hoi, ``Blip-diffusion: Pre-trained subject
  representation for controllable text-to-image generation and editing,''
  \emph{arXiv preprint arXiv:2305.14720}, 2023.

\bibitem{gringoli2019free}
F.~Gringoli, M.~Schulz, J.~Link, and M.~Hollick, ``Free your {CSI}: A channel
  state information extraction platform for modern wi-fi chipsets,'' in
  \emph{Proceedings of the 13th International Workshop on Wireless Network
  Testbeds, Experimental Evaluation \& Characterization}, 2019, pp. 21--28.

\bibitem{wang2019can}
F.~Wang, S.~Panev, Z.~Dai, J.~Han, and D.~Huang, ``Can {WiFi} estimate person
  pose?'' \emph{arXiv preprint arXiv:1904.00277}, 2019.

\bibitem{haarnoja2018soft}
T.~Haarnoja, A.~Zhou, P.~Abbeel, and S.~Levine, ``Soft actor-critic: Off-policy
  maximum entropy deep reinforcement learning with a stochastic actor,'' in
  \emph{International conference on machine learning}.\hskip 1em plus 0.5em
  minus 0.4em\relax PMLR, 2018, pp. 1861--1870.

\bibitem{schulman2017proximal}
J.~Schulman, F.~Wolski, P.~Dhariwal, A.~Radford, and O.~Klimov, ``Proximal
  policy optimization algorithms,'' \emph{arXiv preprint arXiv:1707.06347},
  2017.

\end{thebibliography}
\end{document}